\documentclass[a4paper,11pt]{article}
\pdfoutput=1 

\usepackage{jcappub} 
\usepackage{float}

\usepackage[T1]{fontenc} 
\usepackage{booktabs} 
\usepackage{adjustbox}
\newcommand*{\figuretitle}[1]{%
    {\centering
    \textbf{#1}
    \par\medskip}
}
\newcommand{\ra}[1]{\renewcommand{\arraystretch}{#1}} 
\def\mean#1{\left< #1 \right>}

\title{Angular correlation between IceCube high-energy starting events and starburst sources}

\author{Reetanjali Moharana} \author{and Soebur Razzaque}
\affiliation{Department of Physics, University of Johannesburg,\\ P. O.
  Box 524, Auckland Park 2006, South Africa}

\emailAdd{reetanjalim@uj.ac.za}
\emailAdd{srazzaque@uj.ac.za}

\abstract{
Starburst galaxies and star-forming regions in the Milkyway, with high rate of supernova activities, are candidate sources of high-energy neutrinos. Using a gamma-ray selected sample of these sources we perform statistical analysis of their angular correlation with the four-year sample of high-energy starting events (HESE), detected by the IceCube Neutrino Observatory. We find that the two samples (starburst galaxies and local star-forming regions) are correlated with cosmic neutrinos at $\sim (2-3)\sigma$ (pre-trial) significance level, when the full HESE sample with deposited energy $\gtrsim 20$~TeV is considered. However when we consider the HESE sample with deposited energy $\gtrsim 60$~TeV, which is almost free of atmospheric neutrino and muon backgrounds, the significance of correlation decreased drastically. We perform a similar study for Galactic sources in the 2nd Catalog of Hard {\em Fermi}-LAT Sources (2FHL, $>50$~ GeV) catalog as well, obtaining $\sim (2-3)\sigma$ (pre-trial) correlation, however the significance of correlation increases with higher cutoff energy in the HESE sample for this case. We also fit available gamma-ray data from these sources using a $pp$ interaction model and calculate expected neutrino fluxes. We find that the expected neutrino fluxes for most of the sources are at least an order of magnitude lower than the fluxes required to produce the HESE neutrinos from these sources. This puts the starburst sources being the origin of the IceCube HESE neutrinos in question.
}

\begin{document}
\maketitle
\flushbottom

\section{Introduction}
\label{sec:intro}

The origin of cosmic neutrinos in the sample of high-energy starting events (HESE) detected by the IceCube Neutrino Observatory is one of the outstanding puzzles in recent years~\cite{Anchordoqui:2013dnh}. The four-year HESE sample contains 54 events, collected within 1347 days of operation from both the Northern and Southern skies, with deposited energy in the detector in the range of 20 TeV to 2 PeV~\cite{fra,kop}. These events correspond to a $6.5\sigma$ detection of cosmic neutrinos above an estimated $9.0^{+8.0}_{-2.2}$ atmospheric neutrino background events and $12.6 \pm 5.1$ atmospheric muon background events~\cite{fra,kop}. 

Galactic and extragalctic origin of the HESE events from astrophysical sources \cite{Joshi:2013aua, Murase:2013rfa, Troitsky:2015cnk, Palladino:2016zoe, Neronov:2016bnp} and from dark matter decay \cite{Esmaili:2013gha, Bhattacharya:2014vwa, Murase:2015gea} have been widely discussed in literature. Poor angular reconstruction of the cascade-type events, which dominates the HESE sample, is the main hindrance to identify the sources. Searches for significant association of sources from the same direction as the IceCube neutrino events have been done, importantly with Galactic origin, by the Galactic center~\cite{Razzaque:2013uoa,Fang:2014uja}, Fermi bubbles~\cite{Razzaque:2013uoa,Lunardini:2013gva} and pulsar wind nebulae~\cite{Padovani:2014bha}. In case of extragalactic sources, IceCube neutrino events have been associated with different kind of blazars~\cite{Padovani:2014bha,Krauss:2014tna,Sahu:2014fua,ree1} and with star-forming galaxies~\cite{Anchordoqui:2014yva,Emig:2015dma}.

In our previous study we investigated possible correlation of the three-year HESE sample of 37 events with Ultra-High Energy Cosmic Rays (UHECRs) with energy $\gtrsim 40$~EeV~\cite{Moharana:2015nxa}. We found a $\sim 2\sigma$ (pre-trial) significance of correlation between the HESE sample and the UHECRs with energy $\gtrsim 100$~EeV. A similar study has also been done in Ref.~\cite{chris}, finding similar or higher significance of correlation between the UHECRs and IceCube neutrino events. These studies are independent of the astrophysical source models and useful to check a common origin of the UHECRs and cosmic neutrinos. Subsequently we found that a number of nearby ($z \lesssim 0.06$) Seyfert galaxies in the {\it Swift}-BAT 70 month X-ray source catalog \cite{Baumgartner:2012qx} are within $3^\circ$ error circles of the $\gtrsim 100$~EeV CR directions, which in turn are correlated with the three-year HESE neutrinos~\cite{Moharana:2015nxa}.

Statistical correlations directly between the astrophysical sources and the HESE neutrinos have also been studied. Correlation with different types of blazars, such as high-energy peaked balazars (HBLs)~\cite{Sahu:2014fua}, three-year {\it Fermi} Large Area Telescope (LAT) Active Galactic Nuclei (AGN) Catalog (3LAC)~\cite{ree1}, AGNs in 70 month Fermi-LAT data~\cite{Brown:2015zxa} and Seyfert galaxies~\cite{Emig:2015dma}. No significant correlation was found in these studies. A statistically significant correlation, however, was found between the three-year HESE neutrinos and starburst galaxies and regions~\cite{Emig:2015dma} in the IRAS (Infra-Red Astronomical Satellite) catalog~\cite{Sanders:2003ms}. This is very interesting from astrophysical perspective since supernova remnants, which has a high rate in starburst sources, are widely thought to be the sources of CRs at least up to $\sim 1$~PeV~\cite{Torres:2004hk}. Starburst galaxies have also been proposed as high-energy neutrino sources~\cite{Loeb:2006tw,Tamborra:2014xia}(see, however,~\cite{Murase:2015xka,Bechtol:2015uqb}).  

In this work we study cross correlation between $\gamma$-ray selected starburst galaxies, as well as Galactic sources in the 2nd Fermi Hard source List (2FHL)~\cite{Ackermann:2015uya}, and the four-year HESE sample. We also calculate expected neutrino fluxes from the starburst sources which are correlated with the HESE sample, using $pp$ interaction model to produce observed $\gamma$ rays from them. We describe our data samples in Sec.\ 2 and results of correlation study in Sec.\ 3. We calculate expected neutrino fluxes and compare with required fluxes to explain neutrino data in Sec.\ 4. We discuss our results and conclude in Sec.\ 5.

\section{IceCube HESE neutrino and source samples}
\label{sec:Data}

\begin{figure}[tbp]
\centering 
\includegraphics[width=36pc]{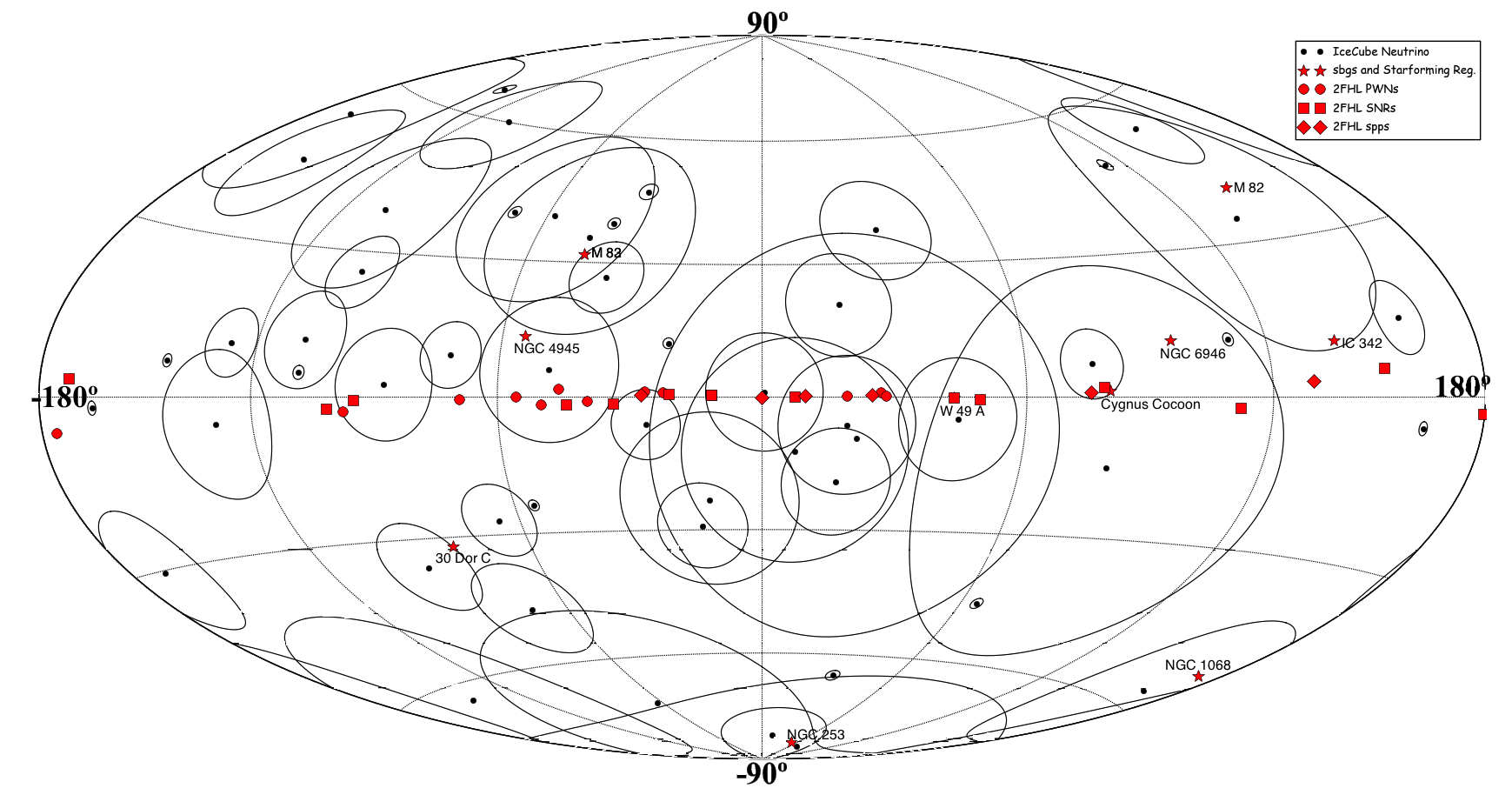}
\caption{\label{skymap} 
Sky Map of the IceCube HESE neutrinos, the shower (39) and track (14) events, in Galactic coordinates. The circular lines the neutrino event directions represent the angular uncertainties. Starburst sources such as starburst galaxies, star-forming regions and the 33 Galactic objects from the 2FHL catalog~\cite{Ackermann:2015uya} are also plotted.}
\end{figure}

We consider the four-year HESE sample of 54 neutrinos with deposited energy in the range $\sim 20$~TeV-2.3~PeV~\cite{kop1} as well as the highest-energy ($\sim 2.6$~PeV) event which happens to be a track event~\cite{kop1}. Two track events (event ID numbers 28 and 32) are coincident hits in the IceTop surface array and are almost certainly a pair of atmospheric muon background events~\cite{aartsen2014observation} and we excluded them from our analysis. Fig.~\ref{skymap} shows sky map of these 53 events in Galactic coordinates with reported angular errors.  

Figure~\ref{skymap} also shows the sources samples we have used to do the analysis. The first sample of sources, named as Sample-I consists of 7 Starburst galaxies (sbgs) and 3 galactic TeV detected star-forming regions. We use the same selection criteria for the 7 starburst galaxies from the IRAS catalog~\cite{Sanders:2003ms} as in Ref.~\cite{Emig:2015dma}. In particular the flux density $S(100~\mu m) \ge 250$~Jy at 100~$\mu m$ was imposed.  Two of the starburst galaxies, namely M~82 and NGC~253, have been detected in TeV $\gamma$ ray~\cite{m82ga,ngc253ga} and appear in the $Fermi$-LAT 3FGL catalog as well.  The other two appear only in the $Fermi$-LAT 3FGL catalog~\cite{2015ApJS..218...23A}, but have not been detected at TeV energies. In another set of source sample, called Sample-II, we included the 4 starburst galaxies from these 7, that are detected by high energy gamma rays in addition to the 3 local starforming regions detected at TeV gamma rays.

We also used 33 Galactic sources that are supernova remnants (SNRs) and pulsar wind nebula (PWNs) from the Second Catalog of Hard $Fermi$-LAT Sources (2FHL) detected between 50~GeV and 2~TeV and called it Sample-III, as these are objects are related to star-forming activities. We have done the analysis for subcategories of these Galactic sources. In particular 12 PWNs called as Sample-IV, 15 SNRs as Sample-V and 6 sources which are either SNR or PWN (SPP) as Sample-VI. Table~\ref{tab:sou} lists all the source samples used for analysis.   

\begin{table}[tbp]\centering
\begin{adjustbox}{width=1\textwidth,center}
\ra{1.4}
\begin{tabular}{|c|c|c|}\toprule
{Source set name}	& {\# of sources} & {Source type}\\ 
\midrule
Sample-I  		   	& 7 		     & 4 IRAS + 3FGL and 3 IRAS  \\
Sample-II  		& 7 		     & 4 IRAS + 3FGL and 3 TeVCAT local Starforming Reg.\\
Sample-III 		   & 33 	     & 2FHL SNRs+PWNs+SPPs \\
Sample-IV 		   & 12 	     & 2FHL PWNs\\
Sample-V  		   & 15 	     & 2FHL SNRs\\
Sample-VI  		   & 6 	     & 2FHL SPPs\\
\bottomrule
\end{tabular}
\end{adjustbox}
\caption{\label{tab:sou} Source sets used to do correlation analysis with neutrino events.}
\end{table}

\section{Cross correlation study}
\label{sec:Analy}

We use the directional information of the neutrino events and sources to map them as unit vectors, ${\hat x}$, on the Galactic sphere and find angular separation between a pair of neutrino and source directions as $\gamma = \cos^{-1} ({\hat x}_{\rm neutrino}\cdot {\hat x}_{\rm source})$~\cite{Virmani:2002xk}. Cross correlation between the neutrino events and sources is based on the number of such pairs observed in data within a given angular error~\cite{Tinyakov:2001nr}.  For our study, we divided twice the reported angular error of each neutrino event ($\delta\theta$) into $M = 20$ concentric rings, each with angular width $\delta\theta/M$. We count the number of pairs $n_j^{\rm data}$ in a given annular ring for $(j-1)2\delta\theta/M \le \gamma < j2\delta\theta/M$, where $j = 1, ... , M.$ This method is different from the minimum $\delta\chi^2$ or nearest neighbor approach used in our previous paper~\cite{Moharana:2015nxa} and allows to explore cross correlation at various angular distances.  

To evaluate the significance of cross correlation between the neutrino events and sources, we generated $N=10^5$ sets of Monte Carlo source distributions by randomizing their positions (more on this later) and count the number of neutrino-source pairs $n_{ij}^{\rm mc}$ in a simulated data set $i = 1, ... , N$ and in a given annular ring $j.$ The p-value is calculated by counting the number of times $n_{ij}^{\rm mc}$ is greater or equal to $n_j^{\rm data}$ in each bin over $N$. The average number of pairs in the Monte Carlo data sets for a given $j$ is ${\bar n}_j^{\rm mc} = \sum_{i=1}^N n_{ij}^{\rm mc}/N$ is the average hits expected for the null distribution. We report the minimum p-value as the significance for that source sample.

We generate Monte Carlo source distributions in two different ways. In the first case, by randomizing the Galactic latitude ($l$) for the sources, keeping their longitude ($b$) fixed. We call this {\it semi-isotropic null} distribution as in Ref.~\cite{Moharana:2015nxa}. In the second case, we randomize $l$ and also $\cos b$ within its allowed range, particularly for Galactic sources we use $|b| \le 10^\circ$ and for other sources we take $|b| \le 90^\circ$. We call this {\it isotropic null} distribution. The agreement between these two null distributions is good.   

The 54 IceCube events contain 21.6 atmospheric neutrino and muon background events. To minimize its effect on the correlation study we have taken two sets of IceCube neutrino events. One with all 53 events and another with neutrino events having energy > 60 TeV, as at higher energy the background is less. The number IceCube neutrino events in the sample having energy > 60 TeV is 33. 

\section{Results and Discussion}
\label{sec:ResD}

\begin{figure}[tbp]
\centering 
\includegraphics[width=0.75\textwidth, trim=0 0 0 0,clip]{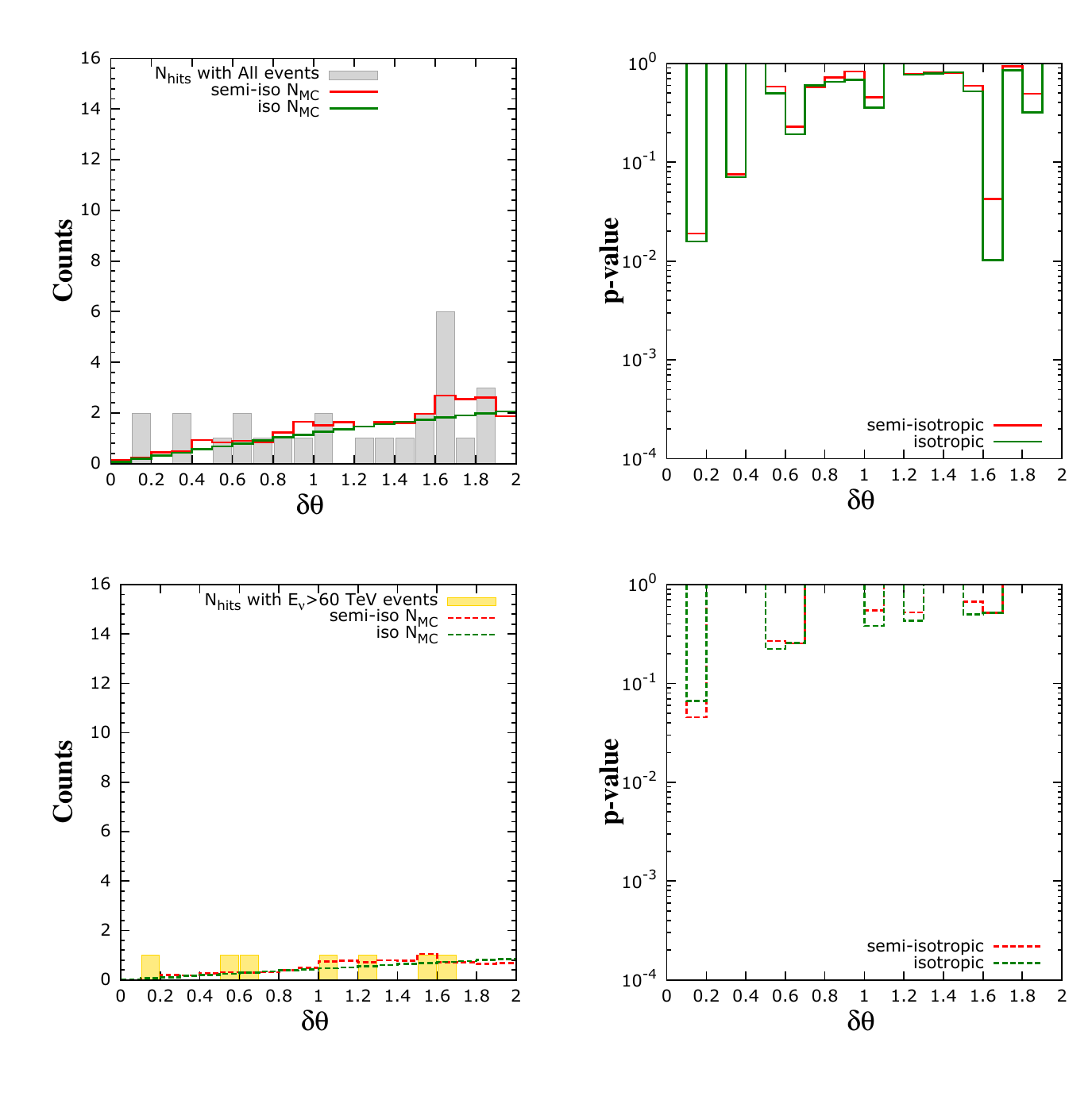}
\caption{\label{fig:sbg}$Top$ : Cross correlation study of the 7 starburst galaxies from the IRAS catalog (Sample-I) with all the 4 Year IceCube neutrino events. Left panel shows, the number of correlation of the IceCube neutrino data with the sources ($n_j^{\rm data}$, with gray histogram) and the average expected number of correlation from Monte Carlo simulation, ${\bar n}_j^{\rm mc}$ with green and red color for $semi$-isotropic and isotropic null distribution respectively, for different bins. Right panel shows the corresponding p-values in each bin for $semi$-isotropic and isotropic null distribution. $Bottom$ : Same as $Top$ but for correlation study with 4 Year IceCube neutrino events with $E_{\nu} > 60$ TeV.}
\end{figure}

\begin{figure}[tbp]
\includegraphics[width=0.75\textwidth, trim=0 0 0 0,clip]{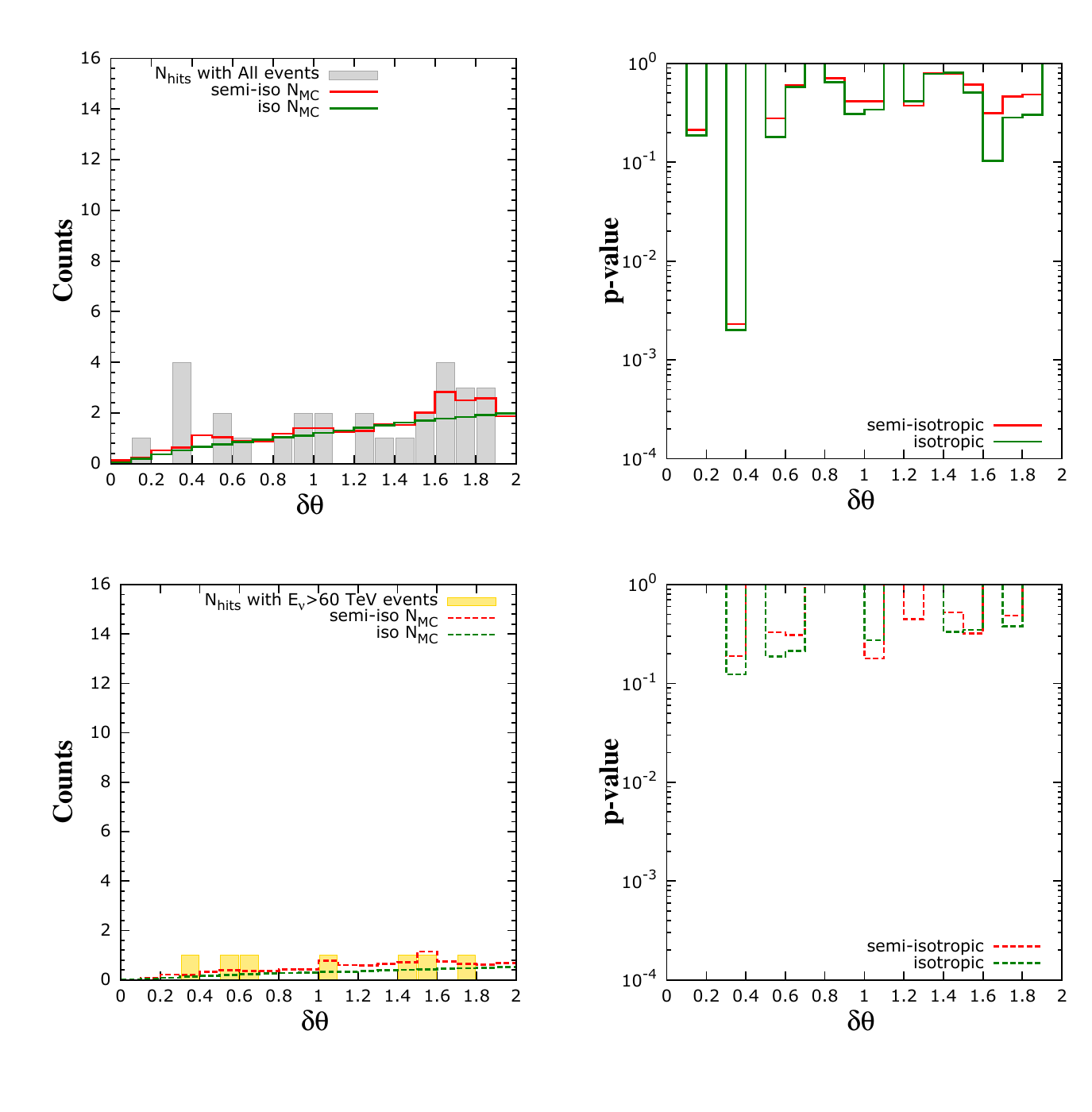}
\caption{\label{fig:sfr} The same as figure~\ref{fig:sbg} for correlation study of IceCube neutrino events with 4 starburst galaxies and 3 local starforming regions (Sample-II) detected in TeV gamma ray energy. }
\end{figure}

\begin{figure}[tbp]
\centering 
\includegraphics[width=0.75\textwidth, trim=0 0 0 0,clip]{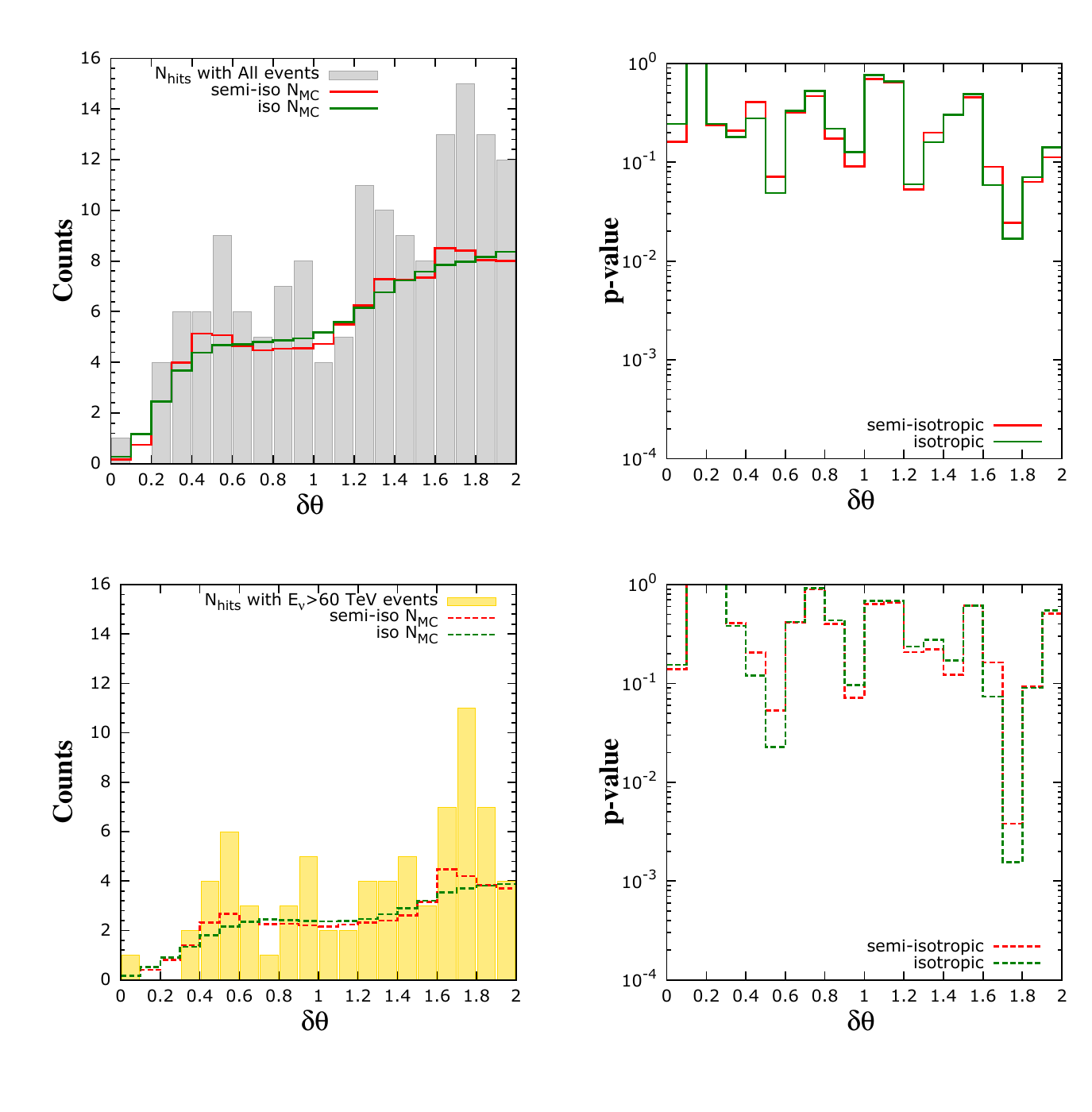}
\caption{\label{fig:gal}Cross correlation study of the 33 Galactic sources from the 2FHL catalog sources (Sample-III) with IceCube neutrino events. The histograms represent the same as in figure \ref{fig:sbg}}
\end{figure}


\begin{figure}[tbp]
\centering 
\includegraphics[width=0.75\textwidth, trim=0 0 0 0,clip]{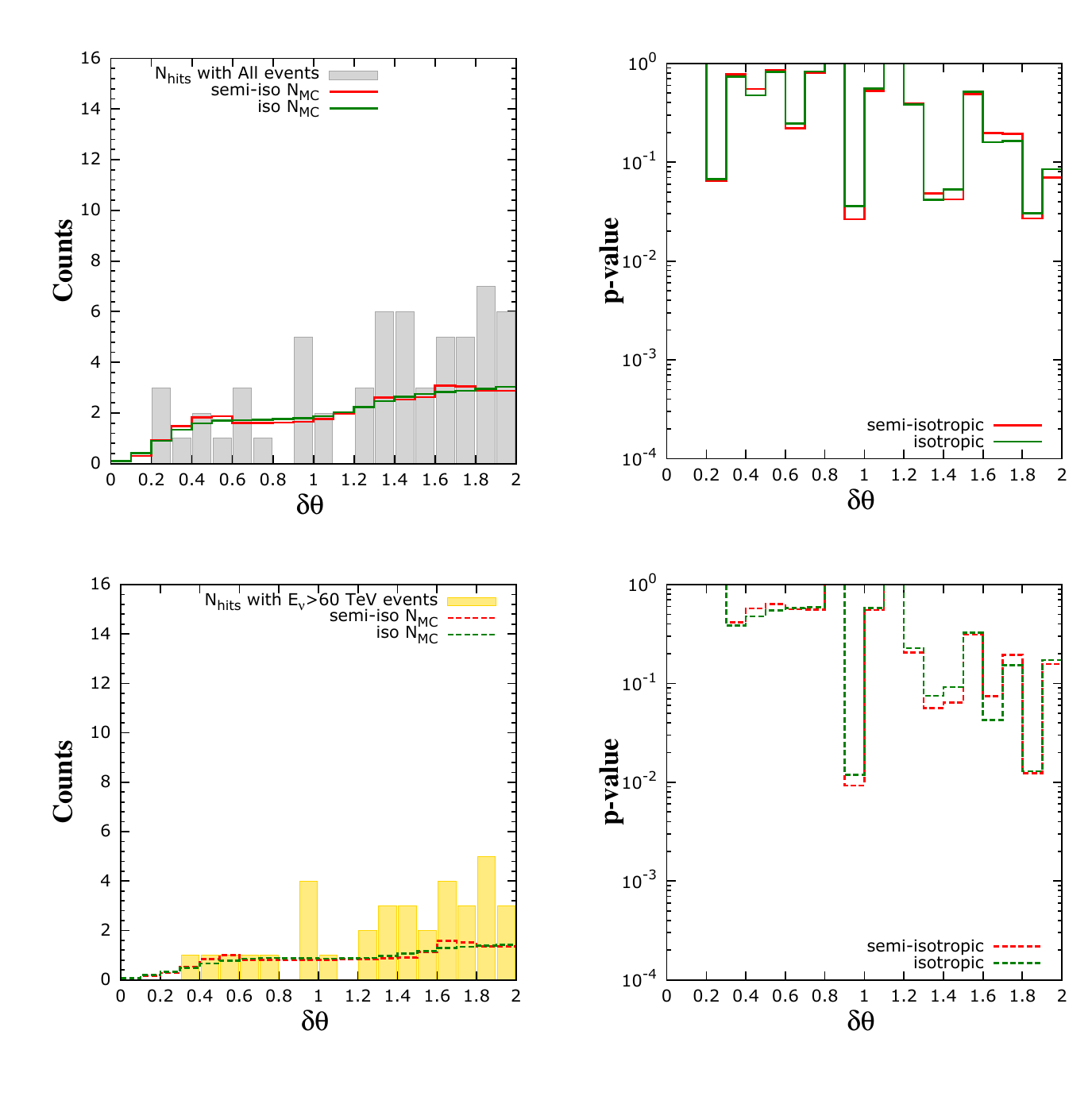}

\caption{\label{fig:2fhl_pwn_dec} Cross correlation study of the 12 2FHL pulsar wind nebulae (Sample-IV) with the IceCube neutrino events following the same color code as in Fig. \ref{fig:sbg}}
\end{figure}

\begin{figure}[tbp]
\centering 
\includegraphics[width=0.75\textwidth, trim=0 0 0 0,clip]{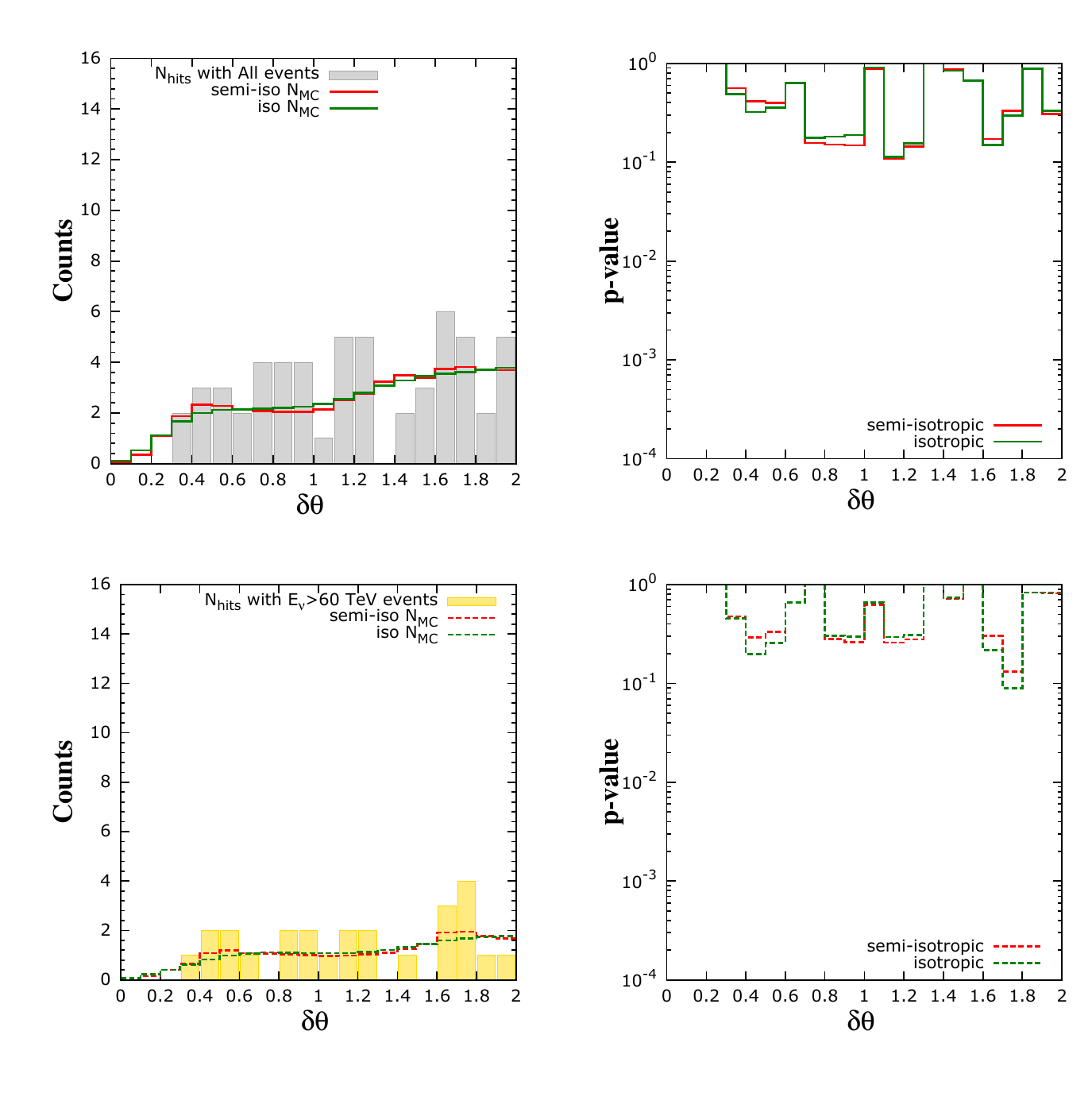}
\caption{\label{fig:2fhl_snr_dec} Cross correlation study of the 15 2FHL super nova remnants (Sample-V) with the IceCube neutrino events following the same color code as in Fig. \ref{fig:sbg}}
\end{figure}

\begin{figure}[tbp]
\centering 
\includegraphics[width=0.75\textwidth, trim=0 0 0 0,clip]{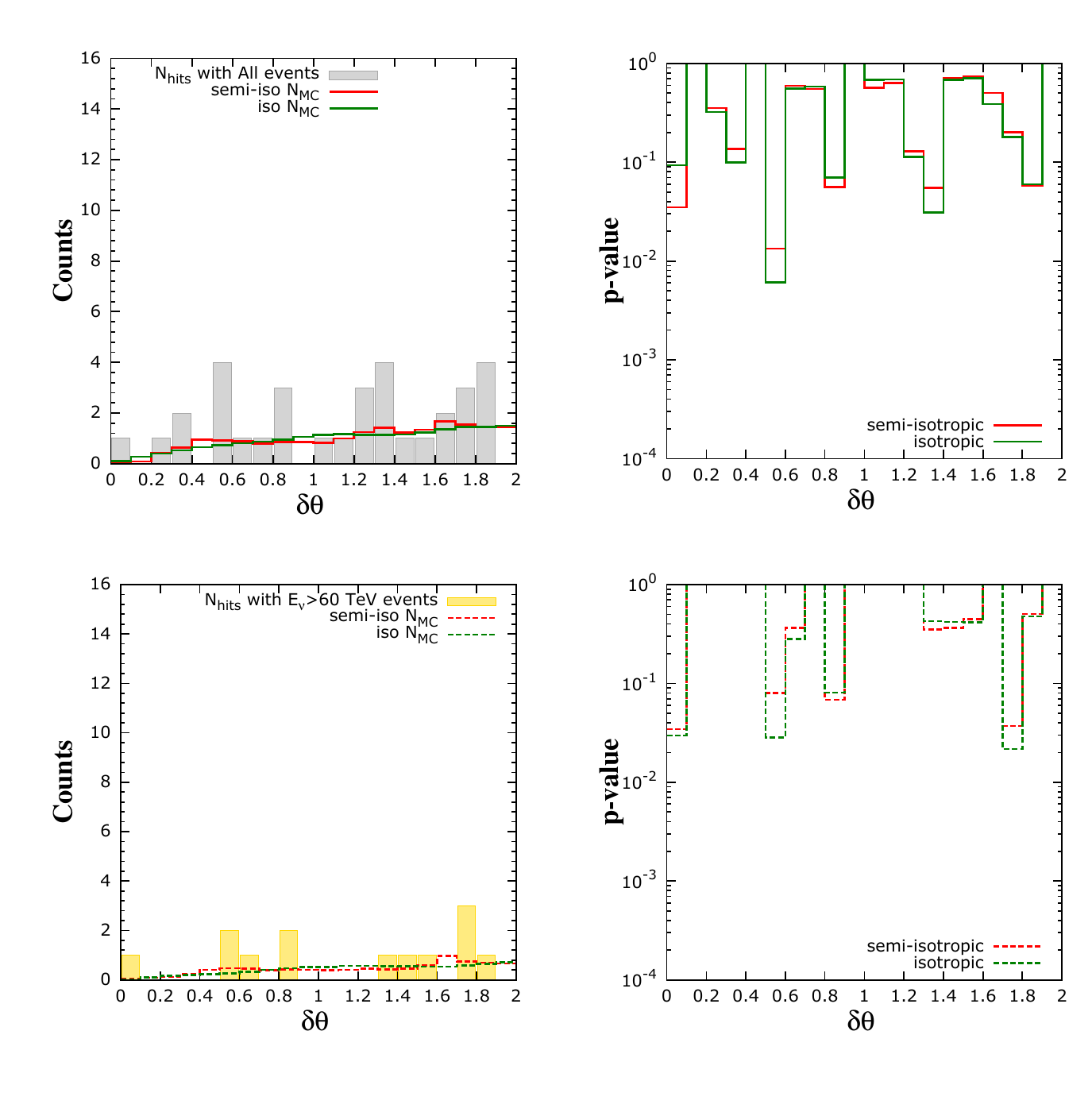}
\caption{\label{fig:2fhl_spp_dec} Cross correlation study of the 6 2FHL SPPs (gamma rays associated with SNRs/PWNs, Sample-VI) with the IceCube neutrino events following the same color code as in Fig. \ref{fig:sbg}}
\end{figure}
The results of cross correlation study on starburst galaxies (Sample-I) with the 53 IceCube neutrino events is shown in figure \ref{fig:sbg}. The $semi$-isotropic and isotropic null distributions are shown with red and green lines, respectively in the left panels. As we have binned over $\delta\theta$ the area swipe in rings increases with increasing the bin number and so does ${\bar n}_j^{\rm mc}$.  The $n_j^{\rm data}$ are shown with gray histogram for the 20 bins.  For Sample-I $n_j^{\rm data}=10$ within the given angular resolution of the IceCube events and the average expected number of hits is 7.7 and 6.1 for $semi$-isotropic and and isotropic null distributions, respectively. In case of the 33 IceCube neutrino events above 60 TeV energy, $n_j^{\rm data}=3$ within the given angular resolution as shown with yellow histogram in the lower panel, while the  ${\bar n}_j^{\rm mc}$ is nearly 2 for the two null distributions. The lowest p-value that we got for Sample-I is nearly 0.01 for the case of all neutrino events in the bin $(1.6 - 1.7)\delta\theta$, for isotropic null distribution. However for the correlation study for events with $E_{\nu}$> 60 TeV there is no significant correlation.

A similar study on the Sample-II is shown in figure~\ref{fig:sfr}. For this sample of starforming region we found $n_j^{\rm data}$ is 11 over an expected count of 8.1 and 6.5 from $semi$-isotropic and isotropic null distributions, respectively in case of all neutrino events within the median angular error reported for IceCube neutrino events. For the events having $E_\nu> 60$ TeV, $n_j^{\rm data}$ is 3 over an expectation of 2.77 and 1.75 from $semi$-isotropic and isotropic null distributions, respectively. The lowest p-value is 0.002 in the bin $(0.3 - 0.4)\delta\theta$ , for the correlation study with all neutrino events for the isotropic null distribution. Like Sample-I we have not found a significant correlation for these sources with the neutrino events above energy 60 TeV.

The cross correlation study of IceCube neutrino events with the 33 Galactic 2FHL sources (Sample-III) is shown in figure~\ref{fig:gal}.  The lowest p-value for the correlation study with all neutrino events is 0.017 in the $(1.6 - 1.7)\delta\theta$ bin, for isotropic null distribution. In this bin the $n_j^{\rm data}$ is 15 while the expected hit is nearly 8 as shown in the upper panel of figure~\ref{fig:gal}. However unlike the first two samples the significance of correlation of neutrino events above energy 60 TeV increased as shown in the lower panel of the figure~\ref{fig:gal}.  The lowest p-value is 0.0016 in the same bin (i.e., $(1.6 - 1.7)\delta\theta$), for the isotropic null distribution.

We have also studied cross correlation of IceCube neutrino events with different type of sources within the 2FHL Galactic sources. Correlation study of IceCube neutrino events with Sample-IV (12 Galactic PWNs listed in 2FHL) sources is shown in figure~\ref{fig:2fhl_pwn_dec}. The lowest p-value for all neutrino events is 0.025 for semi-isotropic null distribution in the $(0.9-1.0)\delta\theta$ bin shown in the upper panel. While for events above 60 TeV energy the minimum p-value is 0.009 in the same bin, shown in the lower panel.

Similarly the correlation study with 15 SNR sources, Sample-V is shown in figure~\ref{fig:2fhl_snr_dec}.  The significance of correlation is lower compared to Sample-IV. From upper panel, the lowest p-value is 0.11 in the bin $(1.0-1.1)\delta\theta$.  And for the lower panel the lowest p-value, 0.089 occurs in the bin $(1.7-1.8)\delta\theta$.

For the spp sources from the 2FHL catalog (Sample-VI) we found a more significant p-value, shown in figure~\ref{fig:2fhl_spp_dec}.  The lowest p-value is 0.006 in the bin $(0.5-0.6)\delta\theta$, from the upper panel. And from the lower panel the lowest p-value, 0.022 occurs in the bin $(1.7-1.8)\delta\theta$.

We have listed the results of this study in Table~\ref{tab:res}. The post trial p-value for $semi$-isotropic and isotropic null distributions are listed in column 4 and 7 respectively.  The post trial p-value is computed as, $1-(1-p)^N$. Where p is the smaller pre-trial p-value between all events and events for energy > 60 TeV of IceCube search. And the number of trial $N$ contains, the number of bins (20) taken within the $2\delta\theta$ directional error reported by IceCube and the two samples of IceCube events taken with cut in energy. Although, strictly speaking, an energy cut does not produce two independent neutrino samples but we are cautious here as we first considered events $>60$ TeV and then all events.  This nevertheless gives more conservative results. We found post trial p-value nearly 0.07 only for the Sample-II and III. We have also computed post-trial p-value while combining different samples.  Samples-IV, V and VI are not independent but sub-samples of Sample-III.  The independent samples are Samples-I, II and III, therefore it may be  interesting to compute a global post-trial p-value, which is $0.09-0.17$, for exploring different samples.  The range here depends on whether treating 60 TeV energy cut producing two independent samples or not.

\begin{table}[tbp]\centering
\begin{adjustbox}{width=1.1\textwidth,center}
\ra{1.3}
\begin{tabular}{@{}|p{1.8cm}||p{2.4cm}|p{2.4cm}|p{2.4cm}||p{2.4cm}|p{2.4cm}|p{2.4cm}|@{}}\toprule
{Source sets} & All $\nu$s $semi$-isotropic random p-value & $E_\nu$> 60 TeV $semi$-isotropic random p-value & Post trial p-value $semi$-isotropic random & All $\nu$s isotropic random p-value & $E_\nu$> 60 TeV isotropic random p-value  & Post trial p-value isotropic random 
\\ \midrule
Sample-I    & 1.9 $\times 10^{-2}$   & $0.518            $    & $0.54$                 & $1. \times 10^{-2 }$   & $6.6\times10^{-2} $      & $0.34$\\
Sample-II   & $2.3 \times 10^{-3}$   & $0.177           $     &$ 8.8 \times 10^{-2}$   & $2 \times 10^{-3}$     & $0.123             $     & $7 \times 10^{-2}$\\
Sample-III  & $2.4 \times 10^{-2}$   & $3.8 \times 10^{-3}$   &  $0.141 $              & $1.7 \times 10^{-2}$   & $1.6 \times 10^{-3}$     & $6 \times 10^{-2}$\\
Sample-IV   & $2.65 \times 10^{-2}$  & $9.23 \times 10^{-3}$  & $0.3112$               & $3 \times 10^{-2}$     & $1.2 \times 10^{-2}$     & $0.38$\\
Sample-V    & $0.11$                 & $0.13$                 & $0.99$                 & 0.15                   & $8.95 \times 10^{-2}$    & $0.97$\\
Sample-VI   & $1.33 \times 10^{-2}$  & $3.4 \times 10^{-2}$   & $0.4149$               & $6.1 \times 10^{-3}$   & $2.12 \times 10^{-2}$    & $0.22$\\
\bottomrule
\end{tabular}
\end{adjustbox}
\caption{Results of cross correlation study for different source sampless analyzed.}
\label{tab:res}
\end{table}

\begin{figure}[tbp]
\centering 
\includegraphics[width=0.7\textwidth, trim=0 0 0 0,clip]{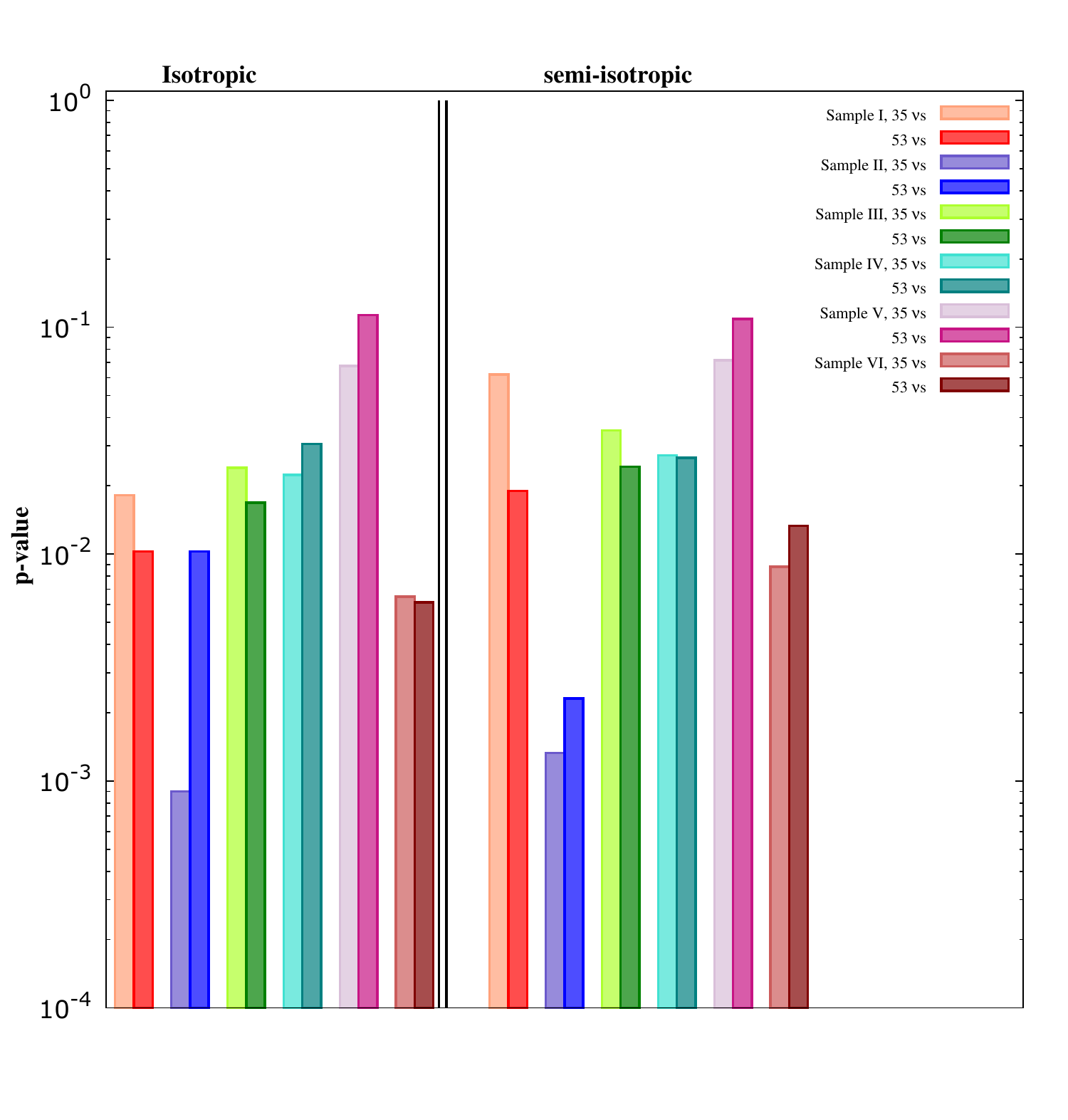}
\includegraphics[width=0.7\textwidth, trim=0 0 0 0,clip]{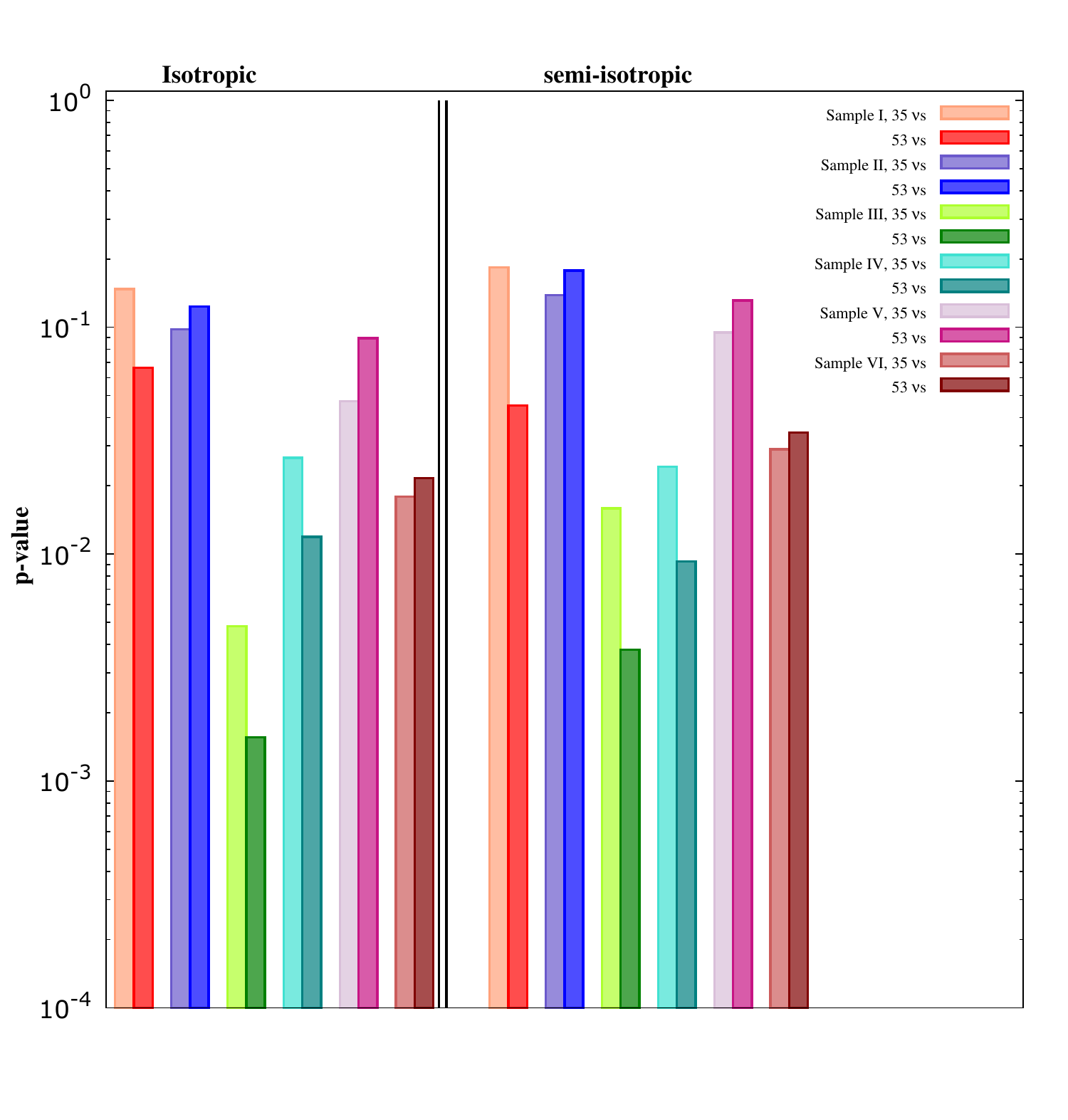}
\caption{\label{compare} 
Comparison plots for all IceCube neutrino events and events with more than 60 TeV energy in case of 3 $yr$ and 4 $yr$ samples are shown in upper and lower panels respectively. }
\end{figure}

A set of sources can be truly associated with the IceCube neutrinos, if the statistic of correlation increases with the increase of neutrino events. To verify whether the starfoming regions are the sources of IceCube neutrino events we have performed the cross correlation study with three year and four year IceCube neutrino sources separately. The results are shown in Fig.~\ref{compare}. The top panel is the result when computed with all the events while the lower panel is for neutrino events with energy more than 60 TeV. The analysis result of the Samples-I and II for 3 $yr$ has also been done in~\cite{Emig:2015dma}, and their result is comparable with our analysis for isotropic null distribution, shown in the top-left panel of figure~\ref{compare}. The faded and darker colors are the minimum p-value occurred in the analysis within the 20 bins of $2\delta\theta$ for 3 year and 4 year data respectively. We see an increase in significance with increasing neutrino events only for the sbg sources and for all the Galactic sources combined.

\section{Star-forming regions with $pp$ interaction}
\label{sec:Cal}
The rate of core-collapse SN generally follows the star-formation rate \cite{Dahlen:2004km}, therefore SNRs in star-forming region is expected to accelerate protons and heavy ions  to energy similar to the knee region ($\sim 10^{15}$ eV) of cosmic ray spectrum. High energy gamma rays and neutrinos are suppose to be produce by the interactions between these accelerated protons and surrounding gas in the environment. The neutrino flux is nearly 2/3 of the gamma ray flux from the $pp$ (proton-proton interaction) channel. Here we have fitted the high energy gamma ray flux produced by the IceCube neutrinos correlated to the star-forming sources, and then estimated the neutrino flux assuming the correlated neutrino event(s) originated from those source(s). In case of pulsar wind nebulae the neutrino production from proton photon interaction can dominate the $pp$ channel. So here we have done the analysis of $pp$ interaction of the sbgs, starforming regions and the spp sources that correlated with the IceCube neutrino events listed in Table~\ref{startab}. The parameters used for $pp$ channel fitting are explained below.  Total number of proton at the source per unit energy interval is,
\begin{equation}
 N_p=N_0 \, E_p^{-\alpha} \, exp{(-E_p/E_0)},
\end{equation}
where $E_0$ is the cutoff energy of protons, $N_0$ is the normalization constant and,

\begin{equation}
 N_0=\frac{F_0}{\int_{{E_p}^{min}} ^ {{E_p}^{max}} {E_p^{1-{\alpha}}  exp{(-E_p/E_0)} dE_p}}.
\end{equation}
Here $F_0$ is the total energy in protons. The photon flux from the $pp$ interaction given as in~\cite{Aharonian:1996},
\begin{equation}
 \frac{dN_\gamma}{dE_\gamma}= \frac{2c\tilde{n}\mean{n_H}}{4\pi D_L^2 K_\pi} \int_{E_{\pi,th}}^\infty {dE_\pi} \frac{\sigma_{pp}(E_c)}{\sqrt{{E_\pi^2}- m_\pi^2}} N_p(E_c).
 \label{gamma}
\end{equation}
Here $\tilde{n}$ is the number of pions produced for the given distribution of function, here we have taken it to be 1, $\mean{n_H}$ is the density of the ambient hydrogen gas, and taken to be 1 $\text{cm}^{-3}$.  The inelastic scattering cross section for $pp$ interaction $\sigma_{pp}$ is taken as in \cite{Kelner:2006tc}, {$E_c=(E_{\pi}/K_{\pi})+m_pc^2$ and $E_{\pi,th} = E_\gamma + m_\pi^2c^4/4E_\gamma$}. $D_L$ is the luminosity distance of the source. We have fitted the high energy gamma ray data available for the individual sources that correlated with the IceCube neutrino events using equation (\ref{gamma}) by varying parameters $E_0$, $\alpha$ and $F_0$. The IceCube neutrino flux for the correlated sources is calculated as in \cite{Lunardini:2011br} for detection time 1374 days and the HESE event effective area.

We have calculated the luminosity of CR protons, $L_{CR}$ from the sources by, $L_{CR}=\int_{E_p^{min}}^{E_p^{max}}{N_p \, E_p \, dE_p}/(t_{pp})$, where $t_{pp}=(\langle n_{H} \rangle\sigma_{pp}c)^{-1}$ is the cooling time of proton. We used $E_{p}^{min}$ corresponding to the pion threshold energy, $E_{\pi,th}$, for $E_\gamma = 70$~MeV and $E_{p}^{max} = 10E_0$.
We have also calculated the gamma-ray luminosity, 
$L_{\gamma}=4\pi D_L^2 \int {E_\gamma} (dN_\gamma / dE_{\gamma}) dE_\gamma$ from fits to the data, for the above mentioned $E_{p}^{min}$ and $E_{p}^{max}$, and listed them in Table \ref{startab}.

\subsection{Sample-I and II}

\begin{figure}[htb]
\centering
    \figuretitle{Sources from Sample-I and II}
  \begin{tabular}{@{}cc@{}}
    \includegraphics[width=.5\textwidth]{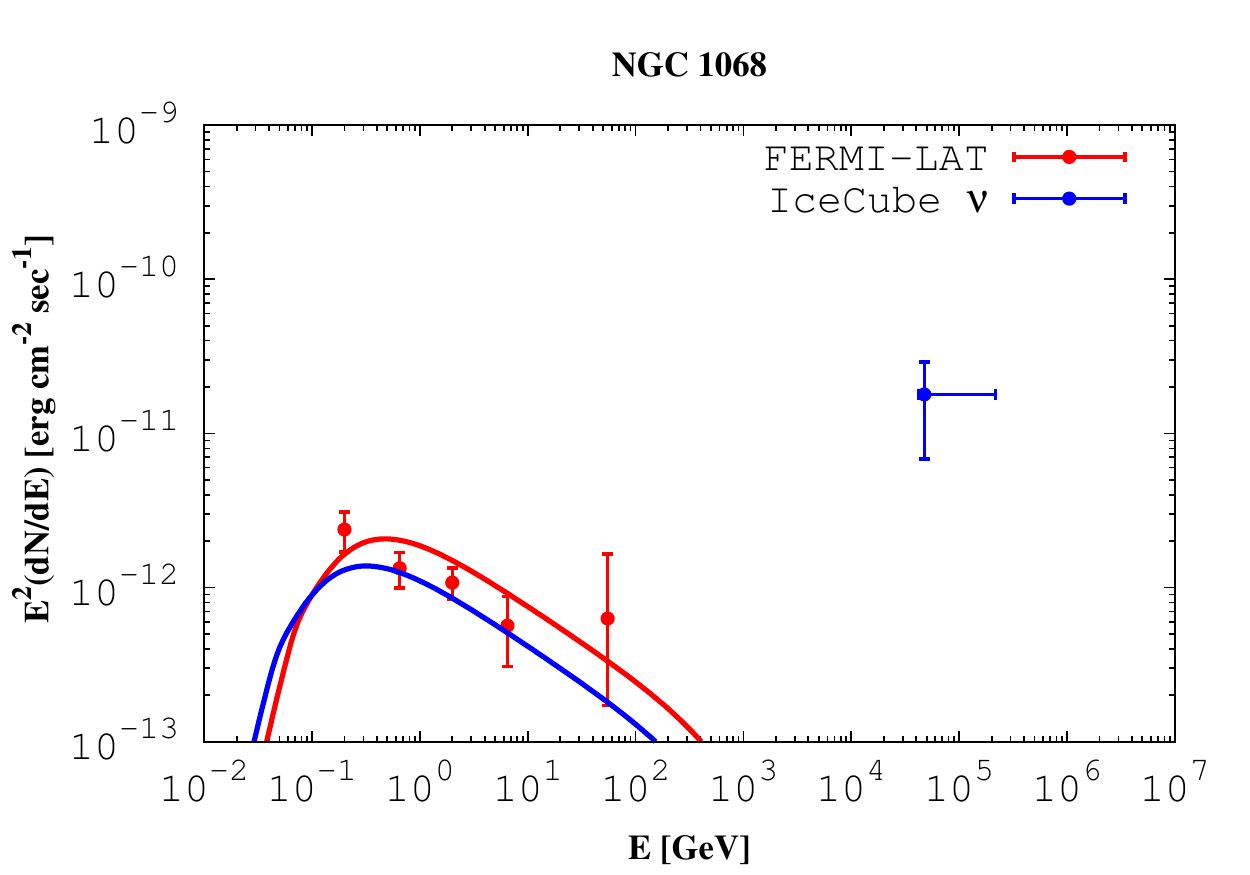}&
    \includegraphics[width=.5\textwidth]{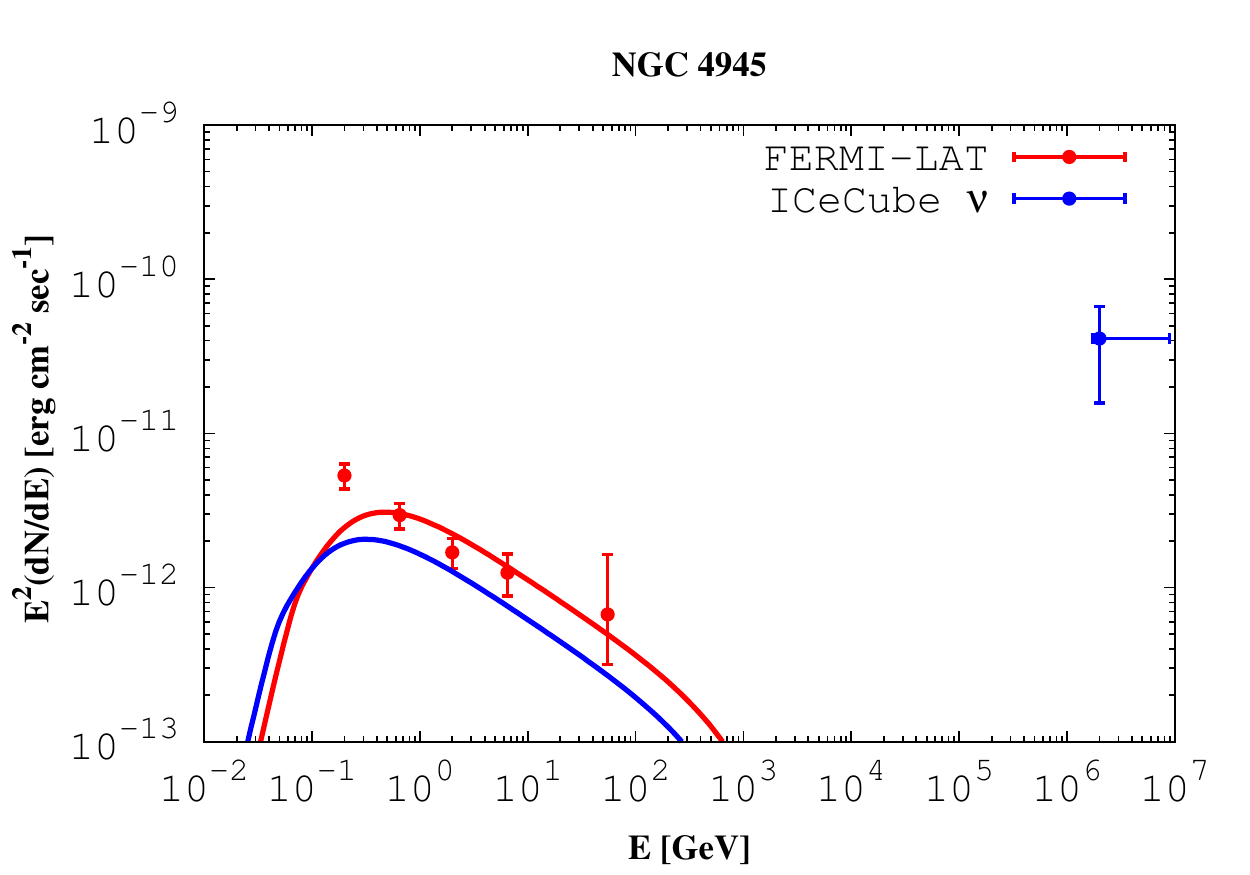} 
   \\
    \includegraphics[width=.5\textwidth]{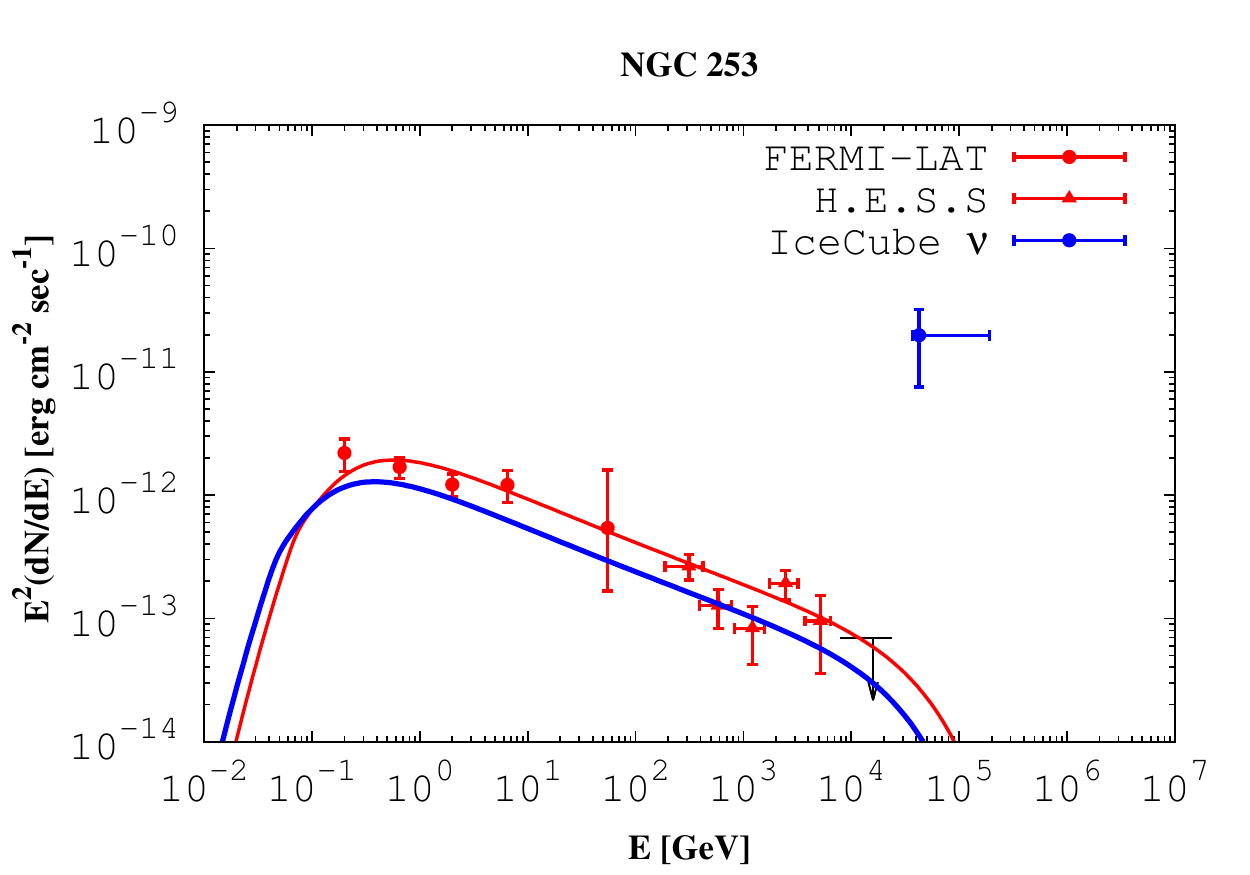} &
    \includegraphics[width=.5\textwidth]{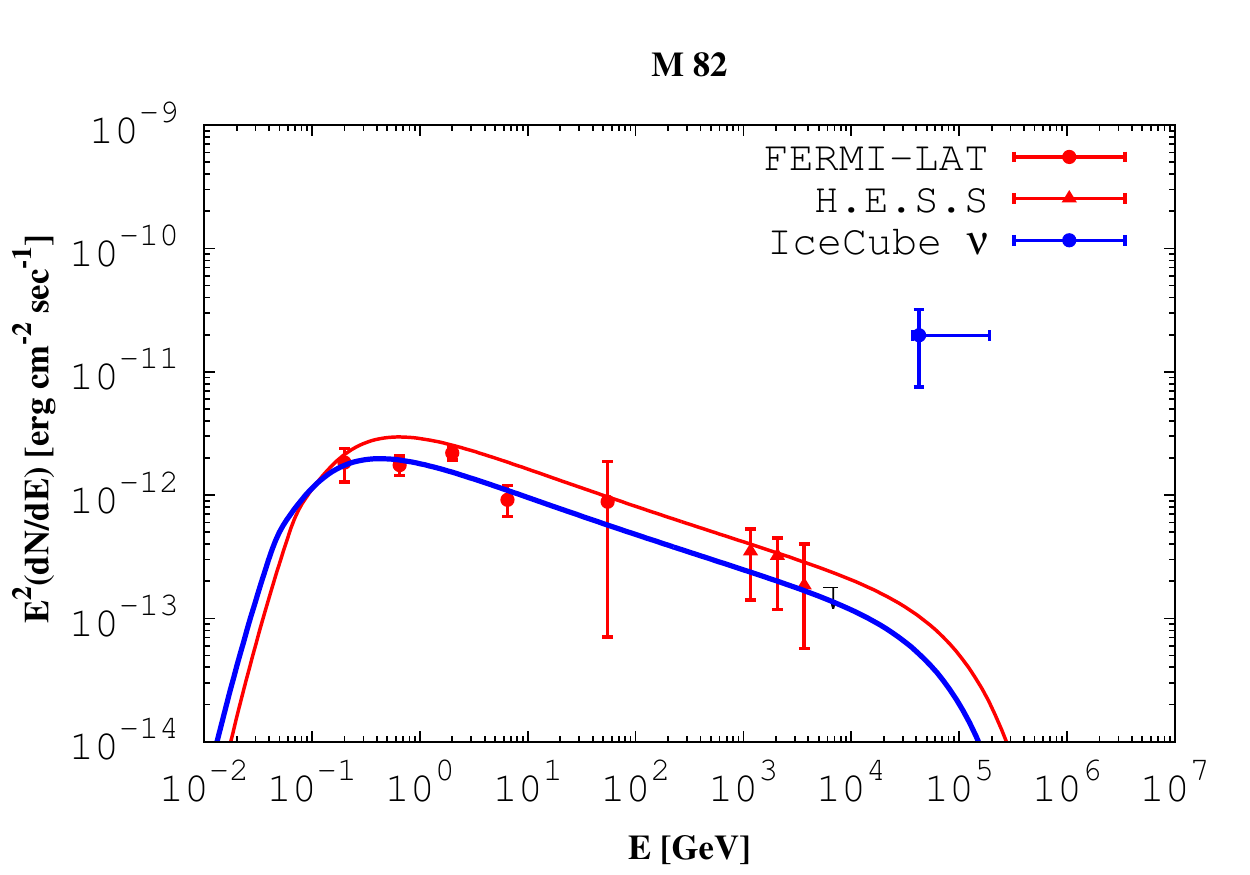}  
   \\
    \includegraphics[width=.5\textwidth]{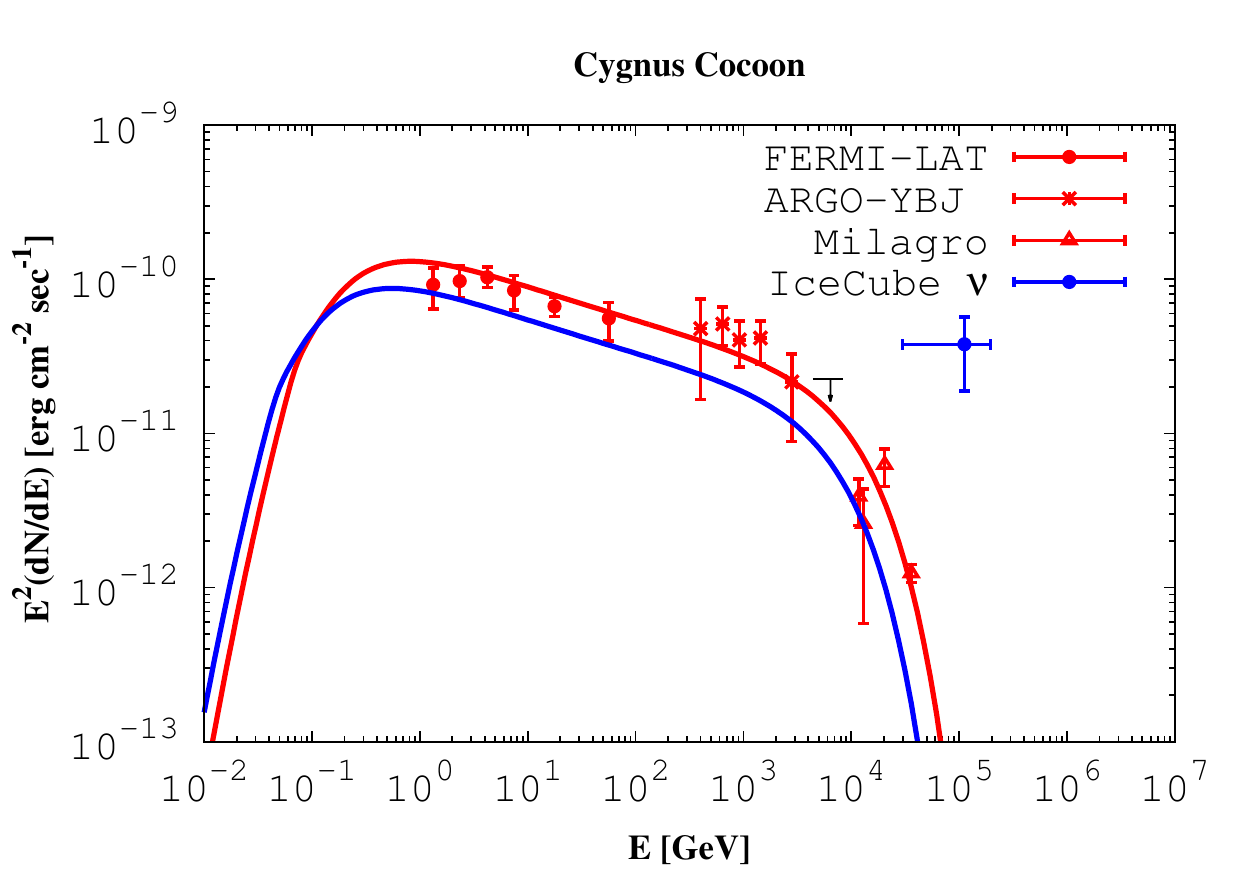} &
    \includegraphics[width=.5\textwidth]{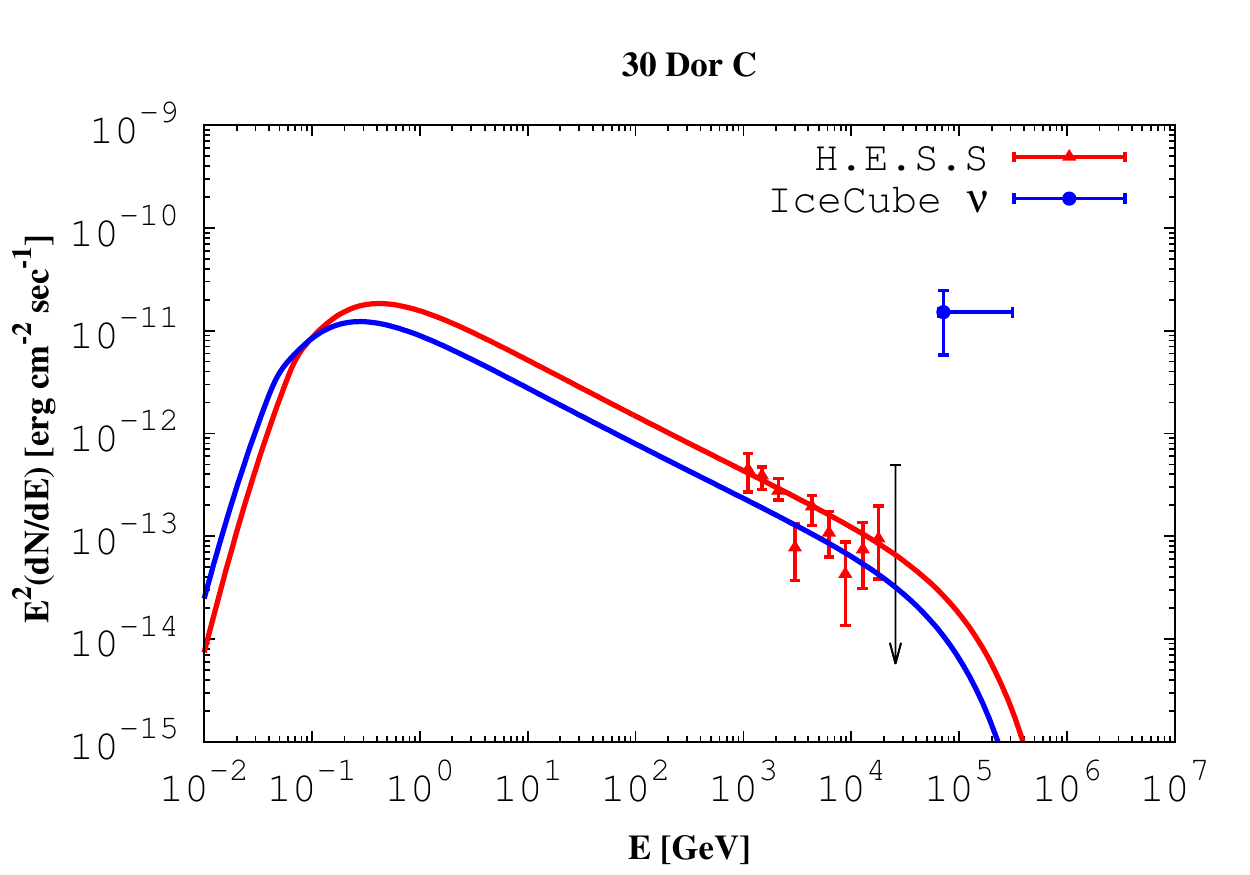} 
  \end{tabular}
 \caption{ \label{saI}The high energy gamma rays detected from sources of Sample I and II, neutrino flux detected from their direction and the $pp$-modeling of the photons. Red circular points represent the data of {\em Fermi}-LAT detected with GeV gamma rays whereas the red triangle represent the H.E.S.S detected events. The IceCube detected neutrino fluxes are shown with blue points. The red line and blue lines show the best fit photon and the corresponding neutrino flux from $pp$ interaction. High energy gamma rays detected from Cygnus Cocoon by  ARGO-YBJ are shown with red star points while by Milagro with open red triangle points as mentioned in \cite{::2014tqa}.}
 \end{figure}

Sample-I and II contains in total 10 sources, 7 are starburst galaxies and 3 are local starforming regions. 
Even though all the 7 sbgs correlated with IceCube neutrino events, only 4 are detected with gamma-rays. So in our $pp$ interaction analysis we have excluded those 3 sources (M 83, IC 342 and NGC 6946). NGC 1068 and NGC 4945 are not detected with TeV events. We have done the fit of gamma rays from $pp$ interaction with the {\em Fermi}-LAT detected events~\cite{2015ApJS..218...23A}, shown in the top panel of figure~\ref{saI}. The middle panel of figure~\ref{saI} shows the Fermi-LAT and H.E.S.S detected gamma rays from NGC 253 (left)~\cite{ngc253ga} and VERITAS detected gamma rays from M 82 (right)~\cite{m82ga}. It also shows the IceCube neutrino flux from their directions as well as the best fit $pp$ interaction fluxes. All the three local starforming regions correlated with one and more IceCube neutrino events. W49 A is in the same location of SNR W49 B. So we have listed it in the SNR sub catalog. Cygnus Cocoon is again in the same position as Gamma cygni for SNR catalog. However we have shown the result here. Cygnus Cocoon is a starforming region in the Milky Way and detected by Fermi-LAT, ARGO-YBJ~\cite{DiSciascio:2015cra} and Milagro~\cite{milcyg}, shown in the left of bottom panel of figure~\ref{saI}. 30 Dor C is a starforming region in the Large Magellanic Cloud and detected by H.E.S.S~\cite{Abramowski:2015rca}, shown in the right of bottom panel of figure~\ref{saI}.

\subsection{Sample-V.}
\begin{figure}[htb]
\centering
    \figuretitle{Sources from Sample-V}
  \begin{tabular}{@{}cc@{}}
    \includegraphics[width=.5\textwidth]{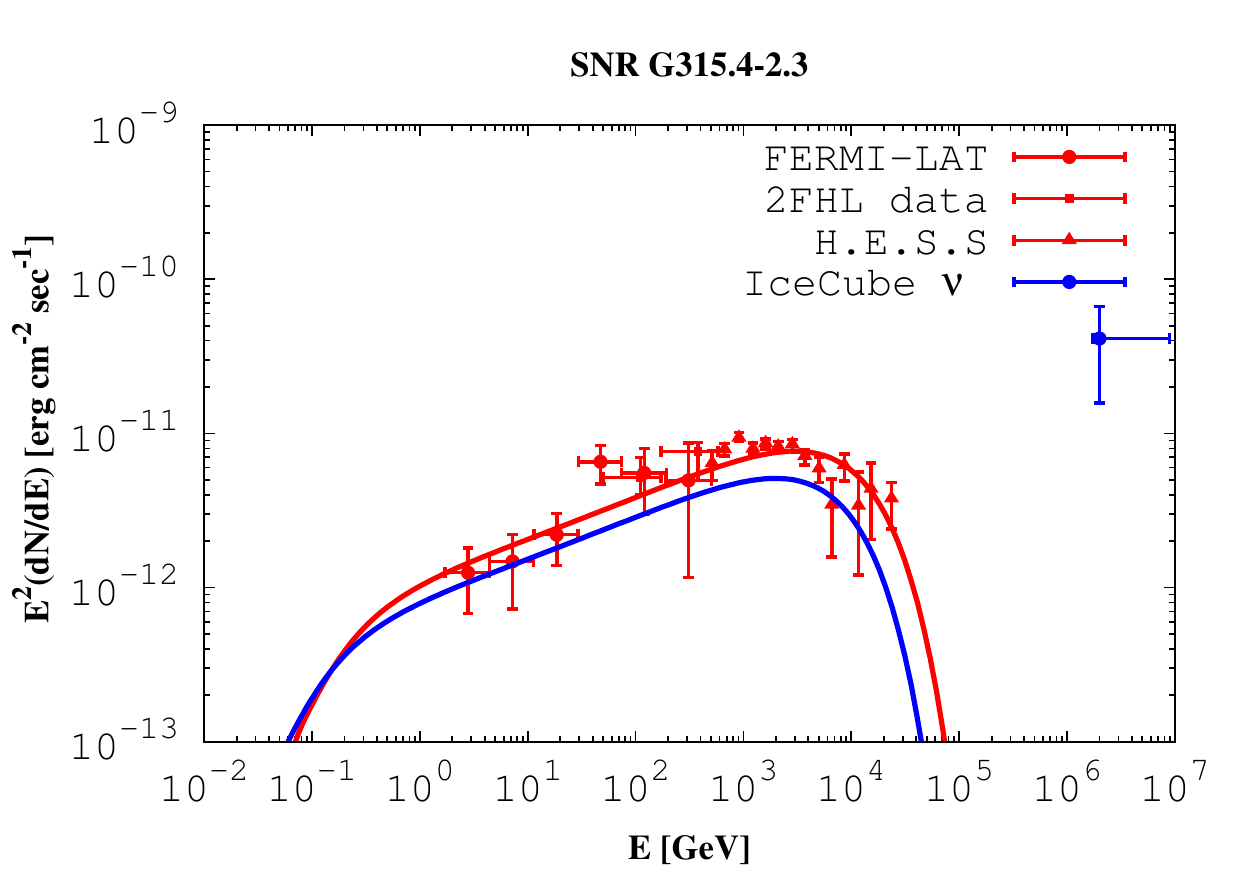} &
    \includegraphics[width=.5\textwidth]{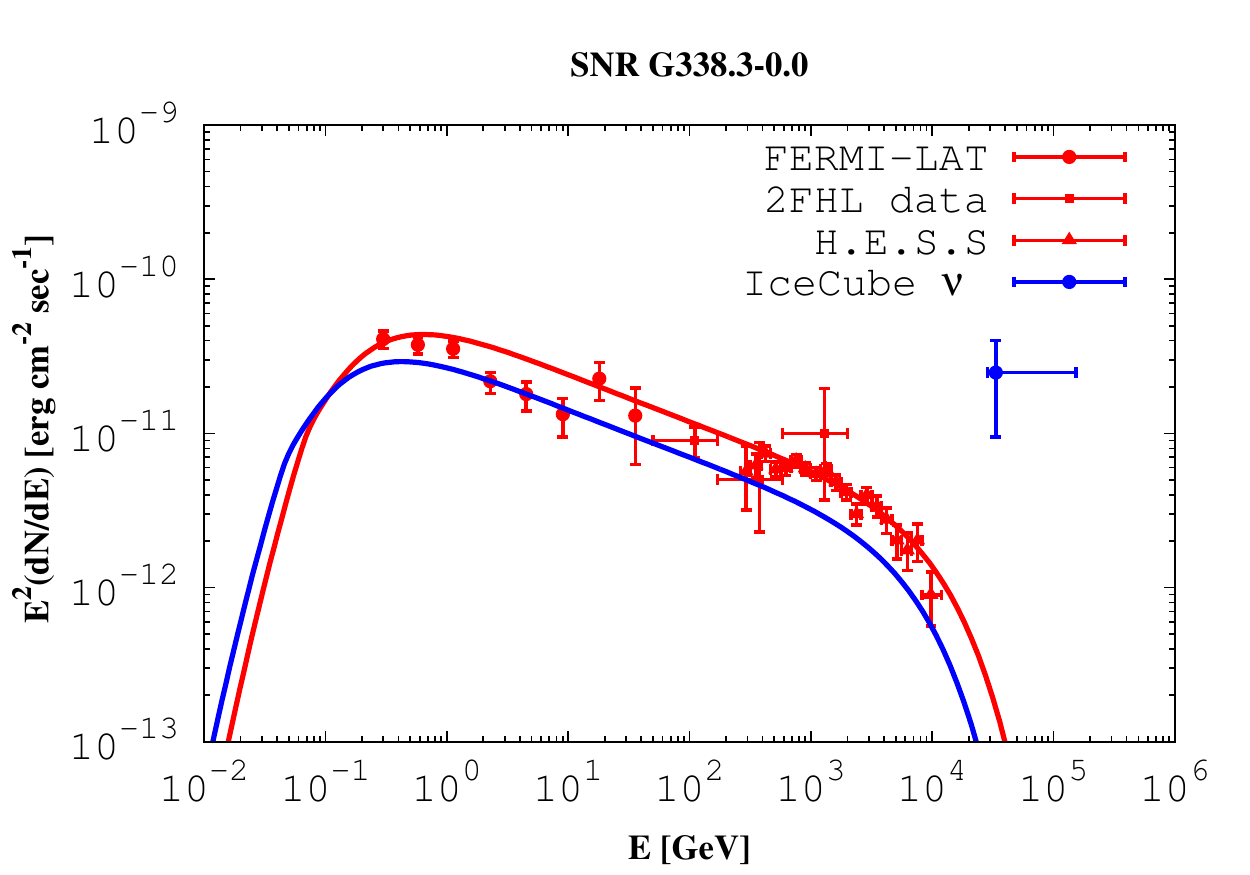} 
   \\
    \includegraphics[width=.5\textwidth]{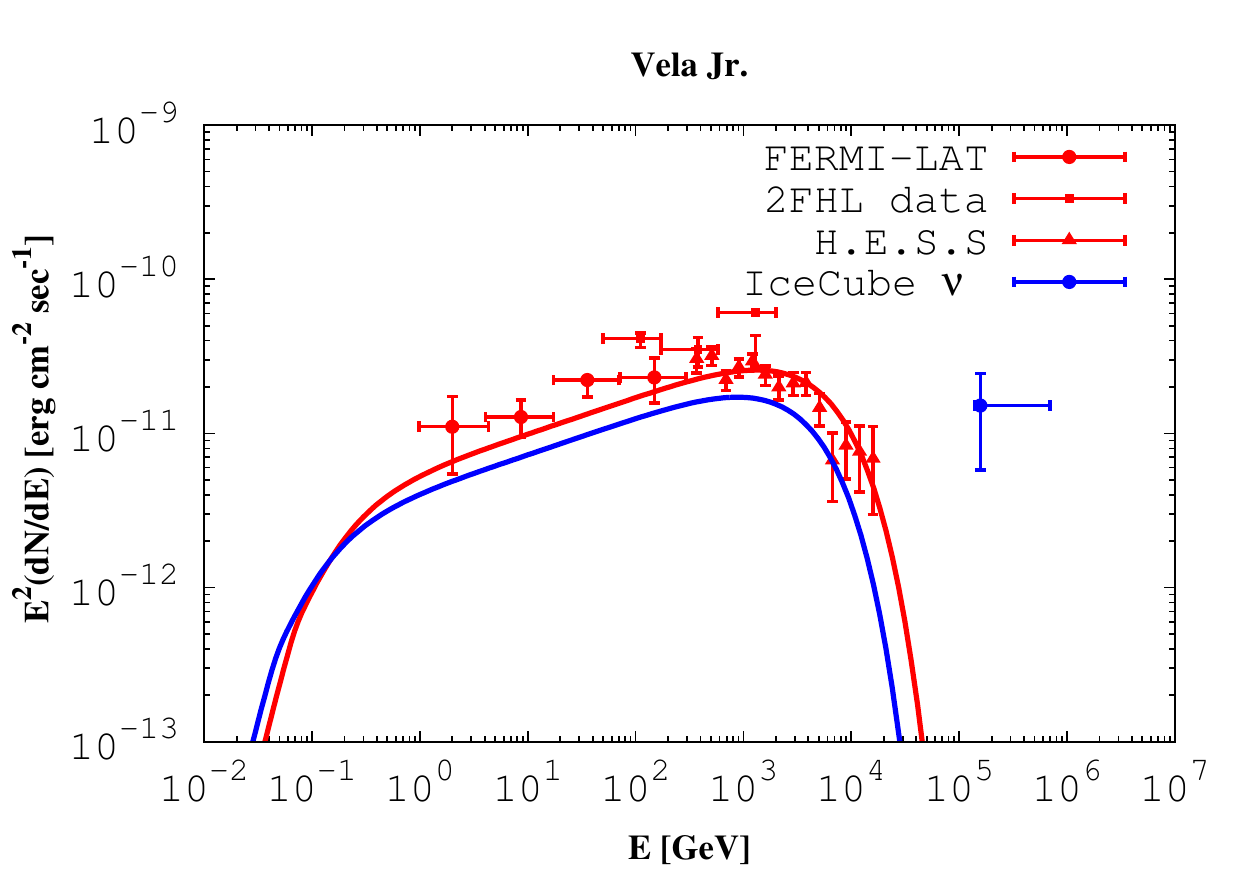} &
    \includegraphics[width=.5\textwidth]{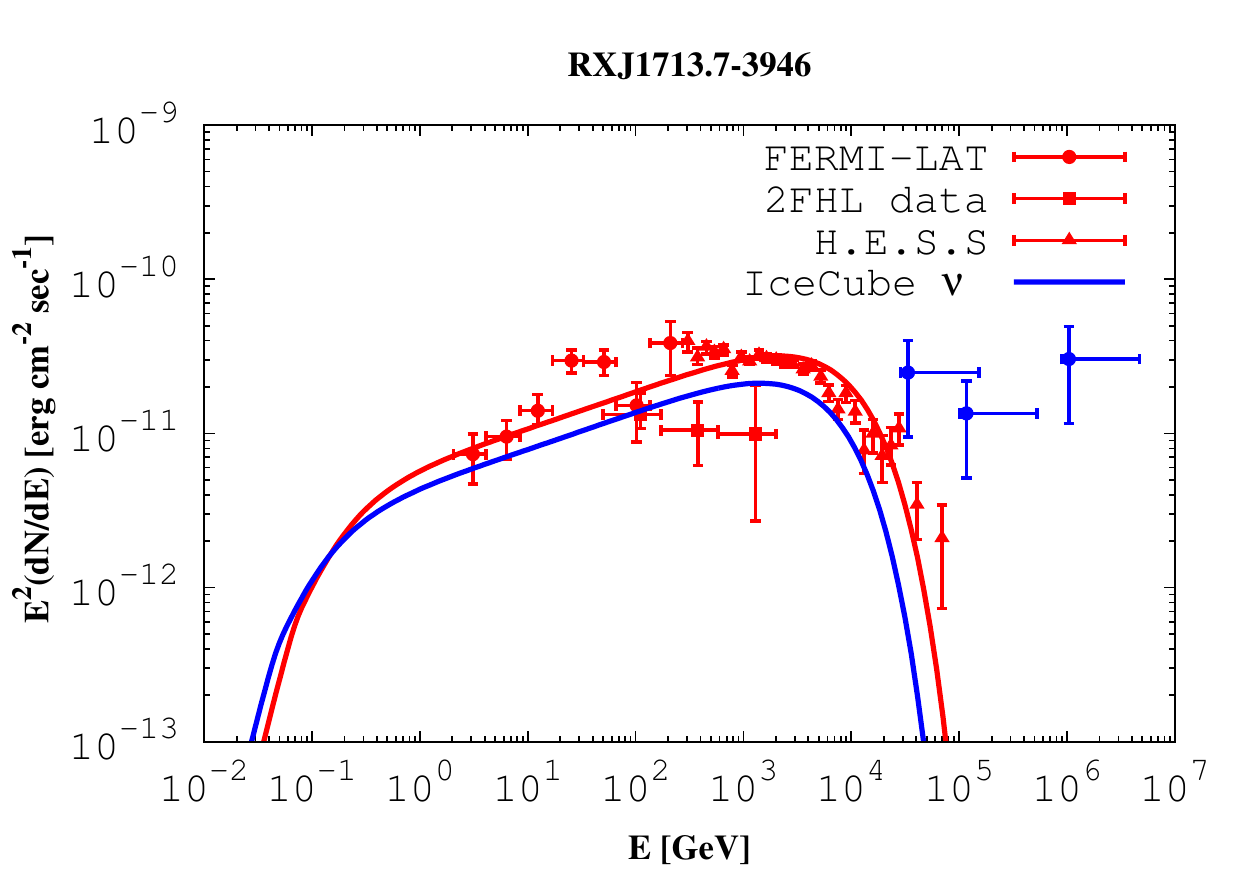}  
   \\
    \includegraphics[width=.5\textwidth]{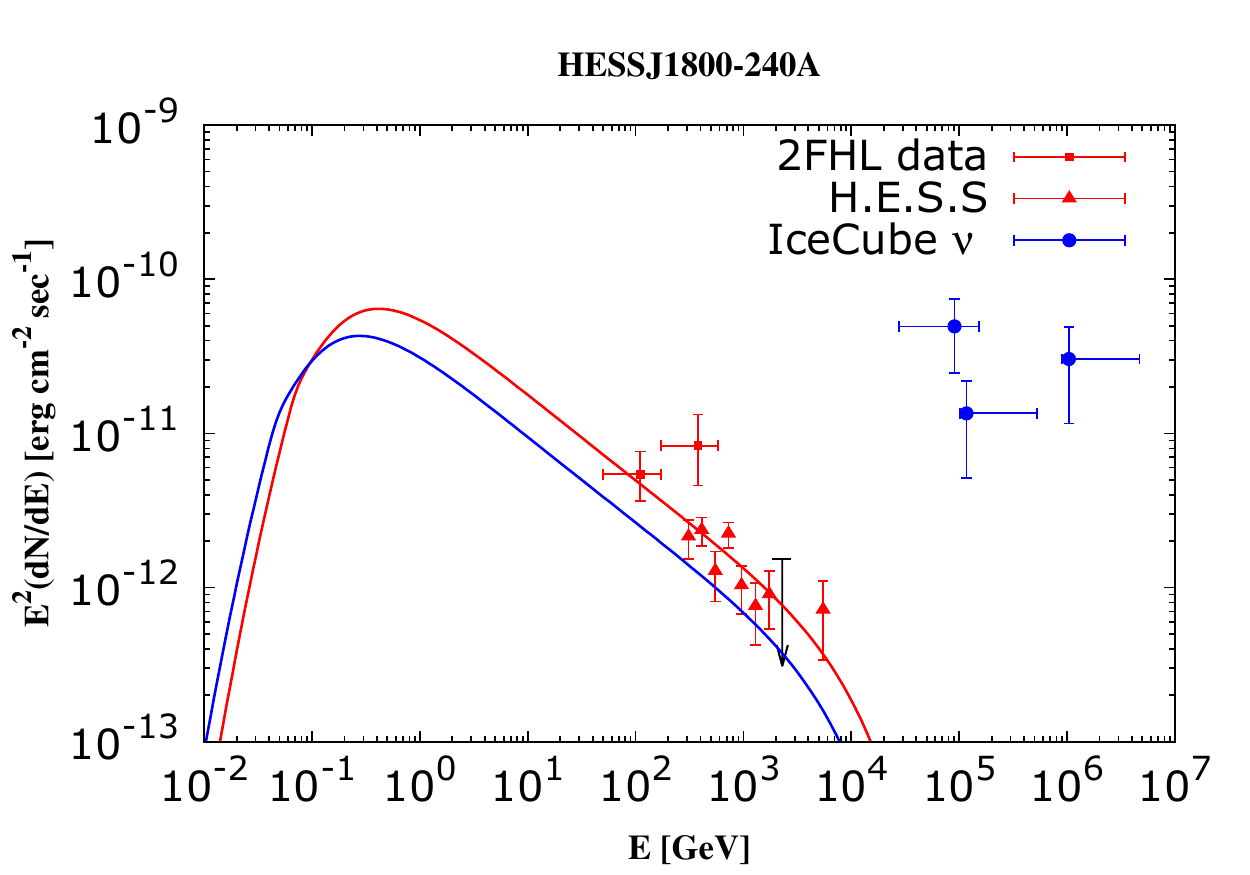} &
    \includegraphics[width=.5\textwidth]{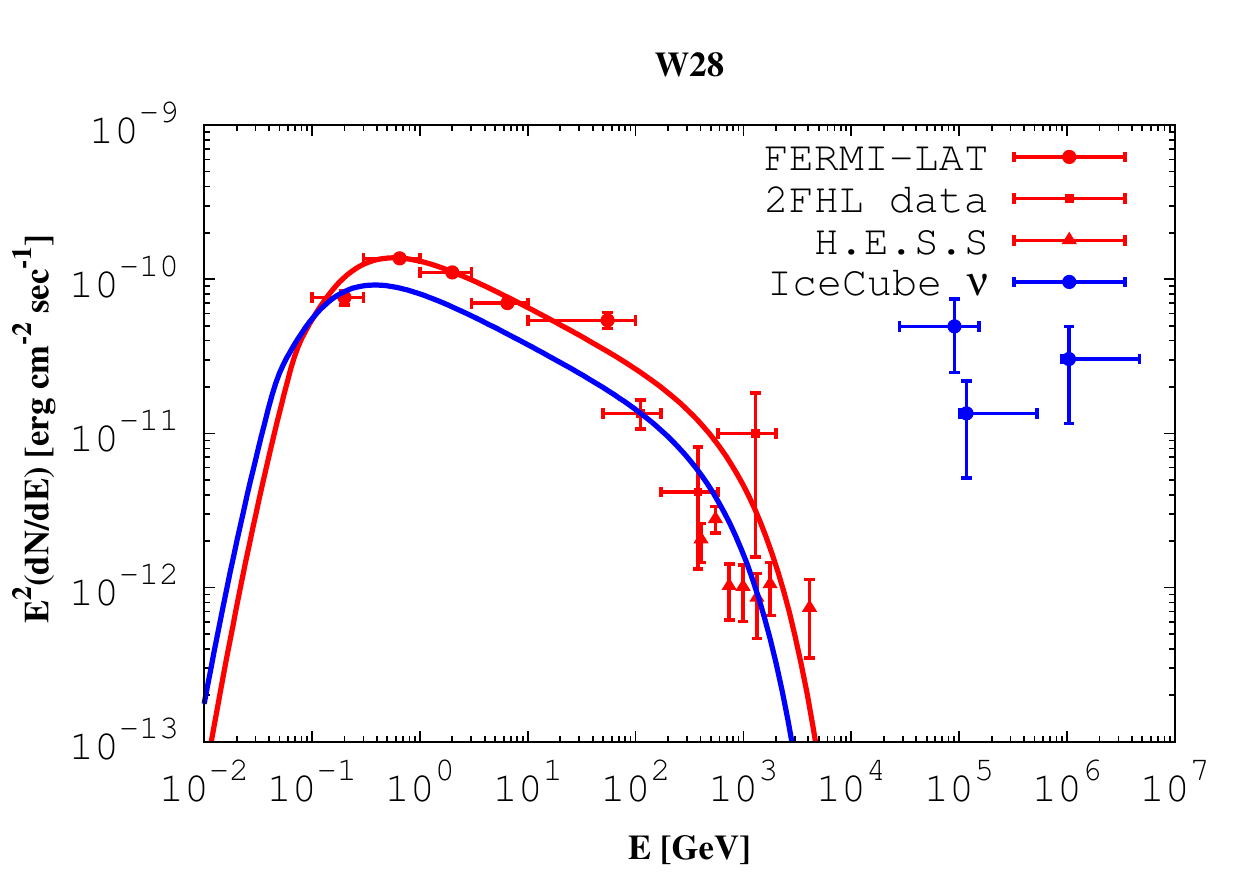} 
  \end{tabular}
 \caption{\label{snr1}Top left panel shows the Fermi-LAT events \cite{fermircw} and H.E.S.S detected gamma rays \cite{hessrcw} from SNR G315.4-2.3. The top right panel shows the Fermi-LAT and H.E.S.S detected events \cite{hesg338} from SNR G338.3-0.0. The middle left panel shows the $Fermi-LAT$ and observed TeV gamma rays by H.E.S.S~\cite{velahess} and the IceCube detected neutrino flux correlated with Vela Jr. The middle right panel shows the Fermi-LAT~\cite{rxfermi} and H.E.S.S~\cite{rxhess} detected gamma rays for RX J1713.7-3946 or 2FHL J1713.5-39. The bottom left panel shows 2FHL data detected by Fermi-LAT and H.E.S.S\cite{Aharonian:2008fp} detected gamma rays for HESS J1800. And the bottom right panel shows the Fermi-LAT \cite{fermiw28} and H.E.S.S ~\cite{Aharonian:2008fp} detected gamma rays for W 28. }
 \end{figure}

\begin{figure}[tbp]
\centering
    \figuretitle{Sources from Sample-V cont...}
  \begin{tabular}{@{}cc@{}}
    \includegraphics[width=.5\textwidth]{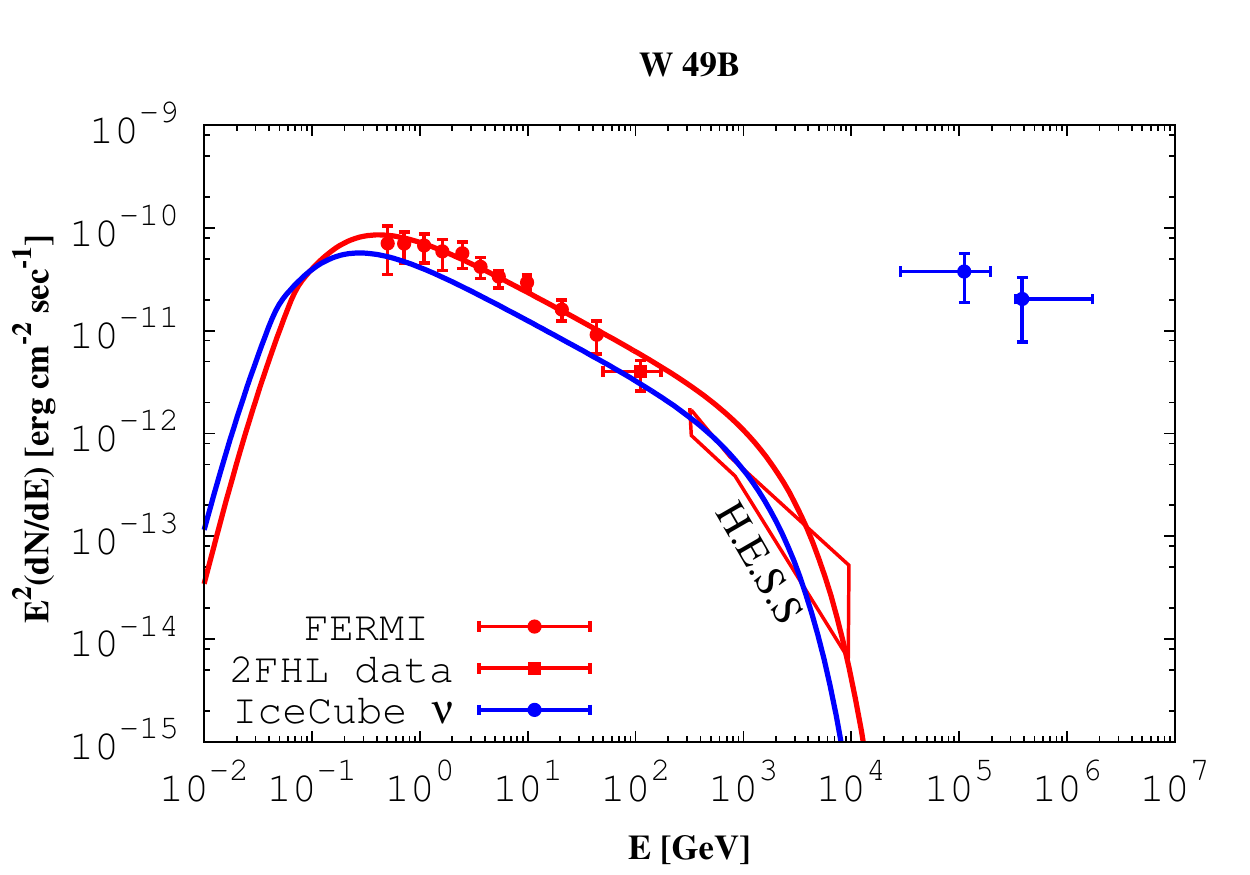} &
    \includegraphics[width=.5\textwidth]{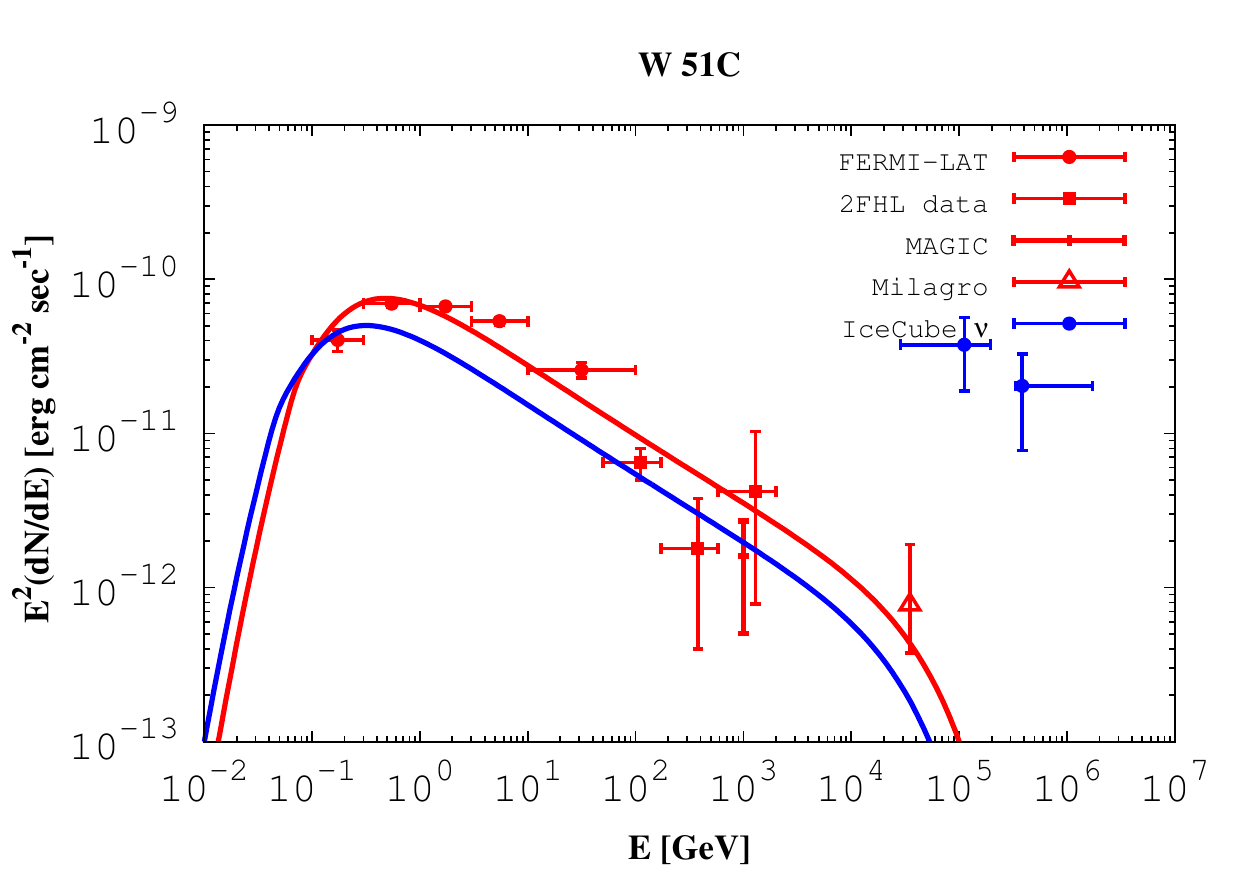} 
   \\
   \multicolumn{2}{c}{\includegraphics[width=.5\textwidth]{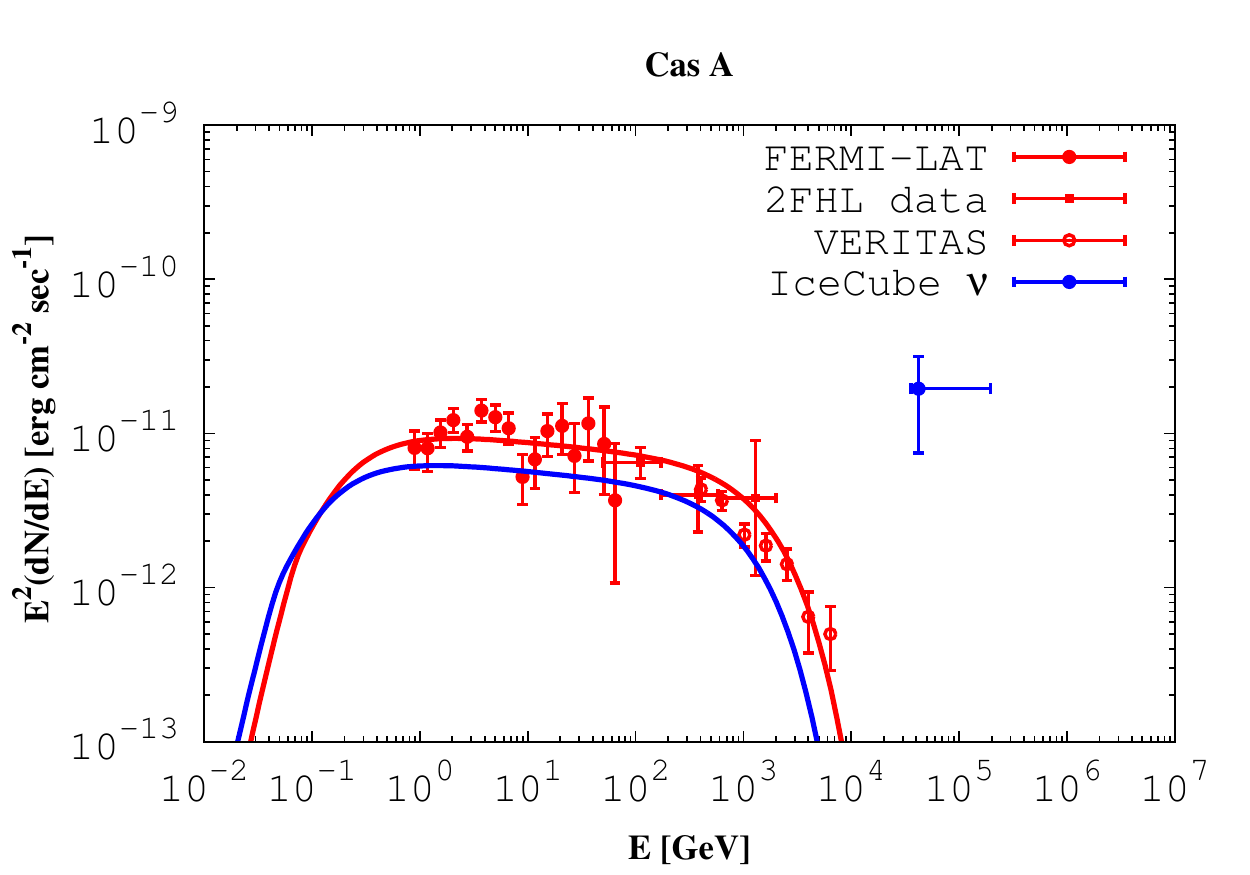}}
   
  \end{tabular}
 \caption{\label{snr2}The top left panel shows Fermi-LAT and H.E.S.S \cite{w49} detected gamma rays from W49. And the top right panel shows the Fermi-LAT and MAGIC \cite{magic_51} and Milagro~\cite{milagro_51} detected gamma rays from W51C with red circle, square and triangle points, respectively. The bottom plot shows the Fermi-LAT and VERITAS~\cite{casa} detected gamma rays from Cas A .}
 \end{figure}
Sample-V contains 15 SNRs from the 2FHL catalog, 10 of these are correlated with the direction of the IceCube neutrino events. The SNRs are surrounded by molecular cloud and high energy cosmic ray protons can interact with these clouds to produce high energy gama rays, like the Fermi-LAT detected gamma rays at the vicinity W28 are explained as in \cite{lara}.  The result of the $pp$ interaction analysis for these sources are shown in figure~\ref{snr1}-\ref{snr2}.

\subsection{Sample-VI}
\begin{figure}[htb]
\centering
    \figuretitle{Sources from Sample-VI}
  \begin{tabular}{@{}cc@{}}
    \includegraphics[width=.5\textwidth]{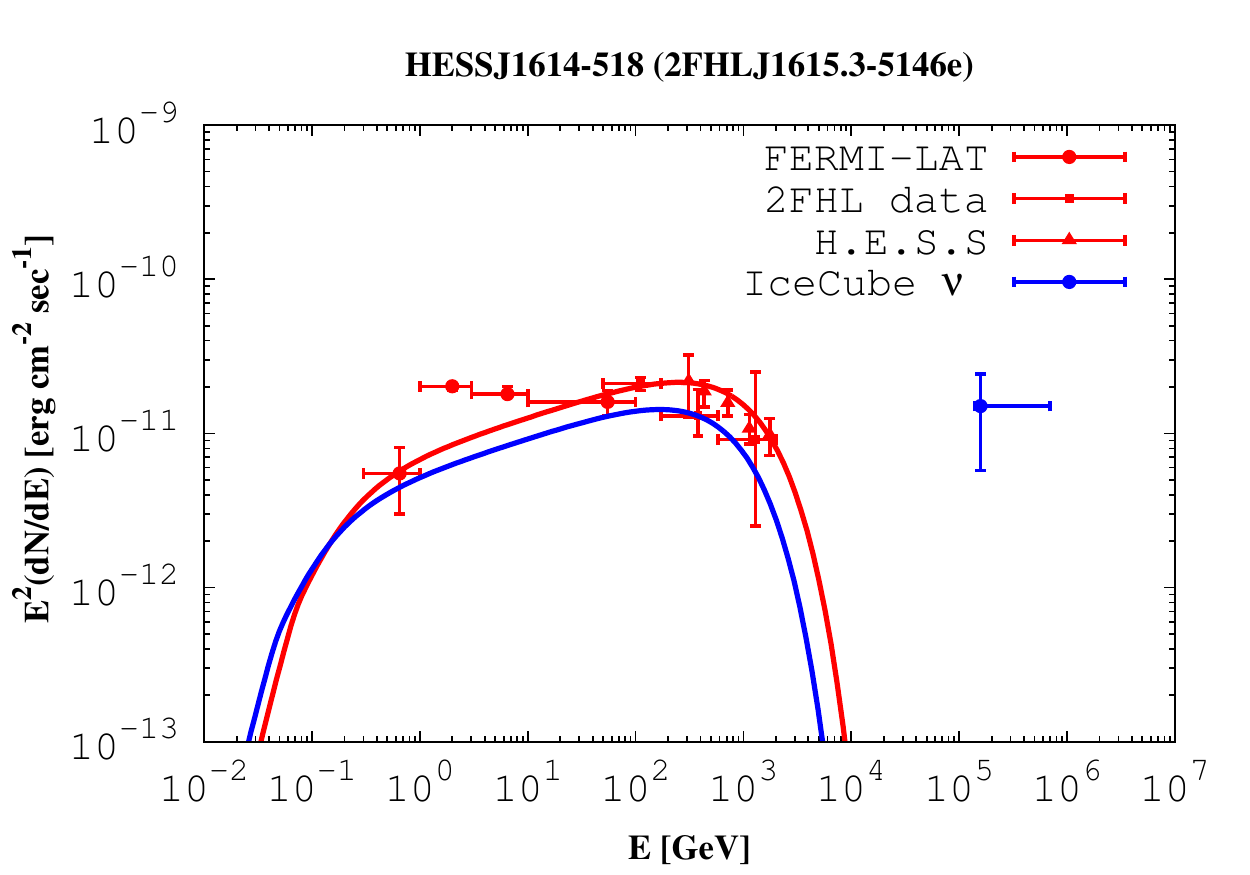} &
    \includegraphics[width=.5\textwidth]{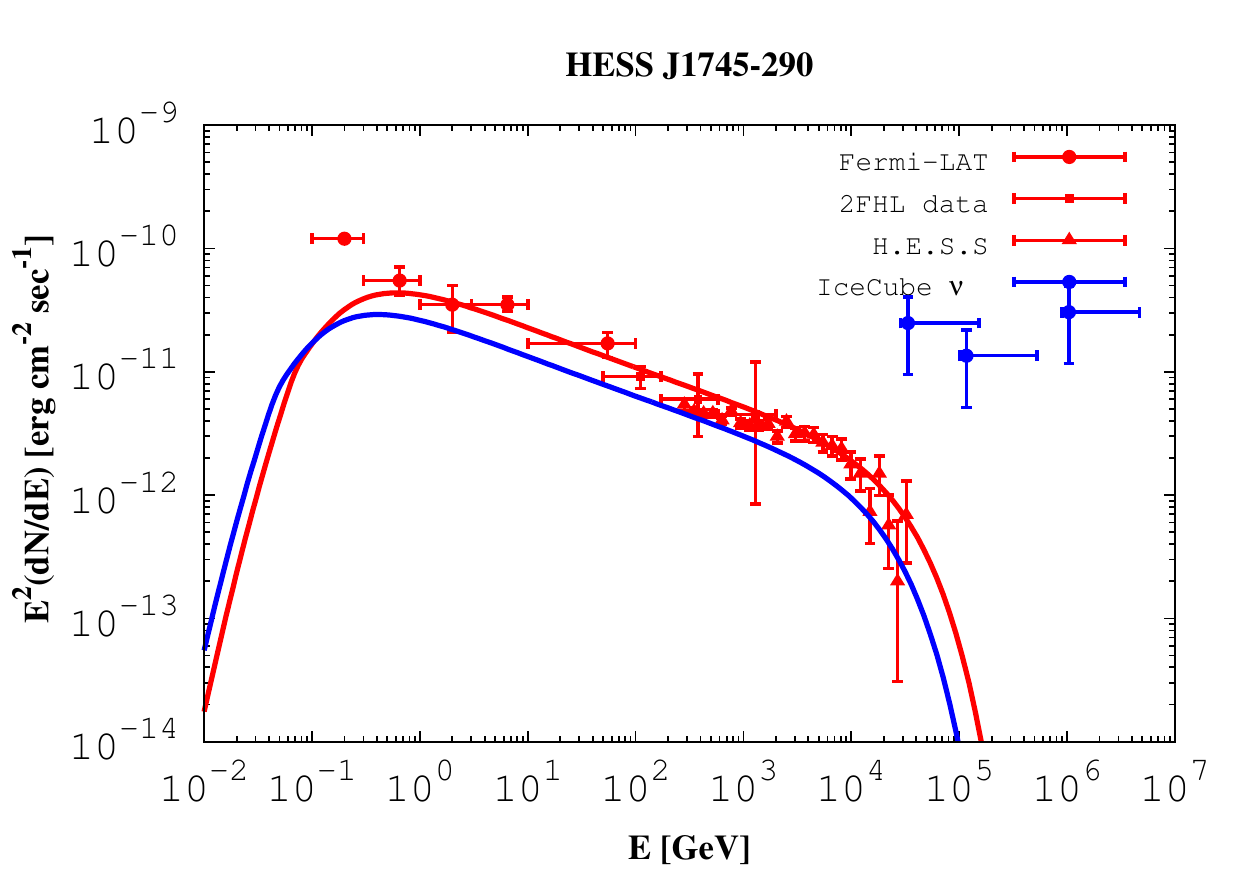} 
   \\
    \includegraphics[width=.5\textwidth]{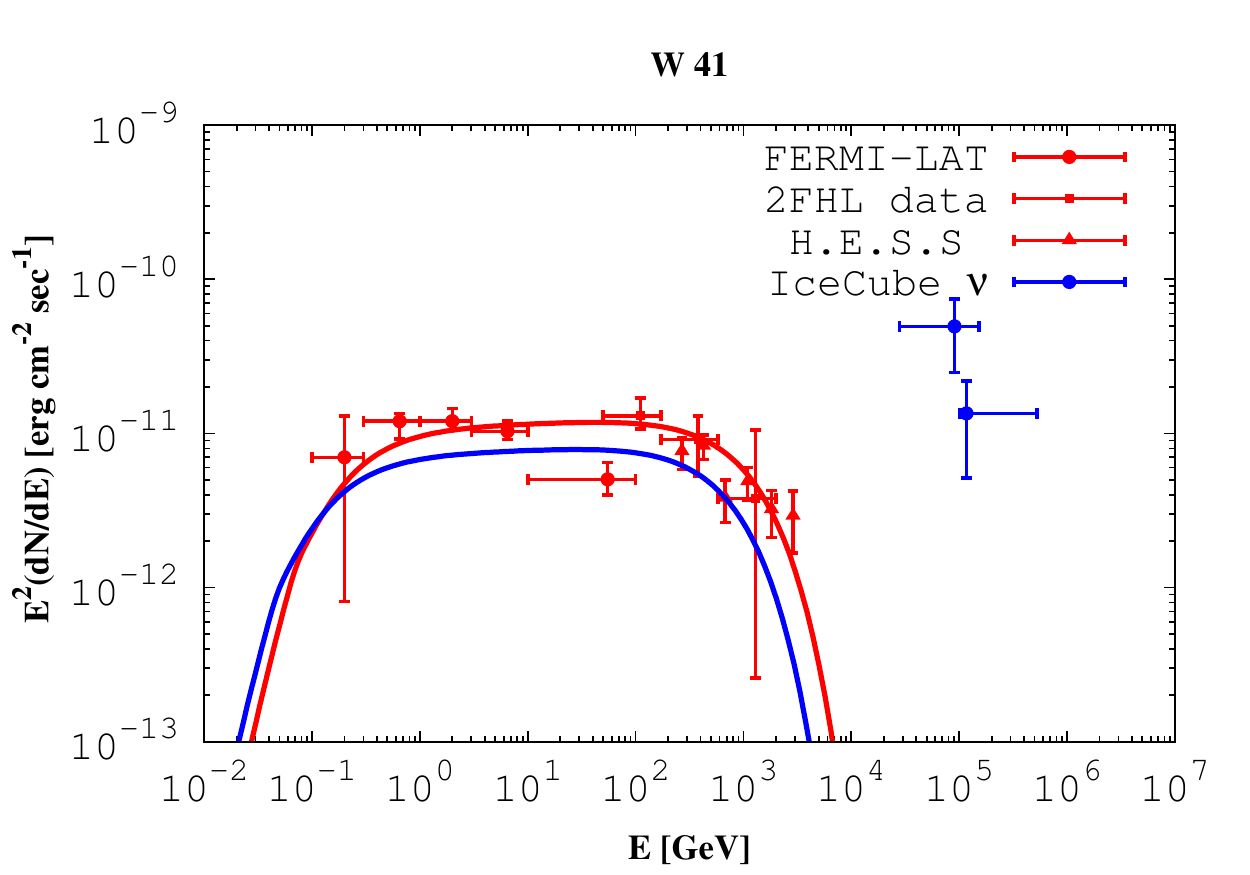} &
    \includegraphics[width=.5\textwidth]{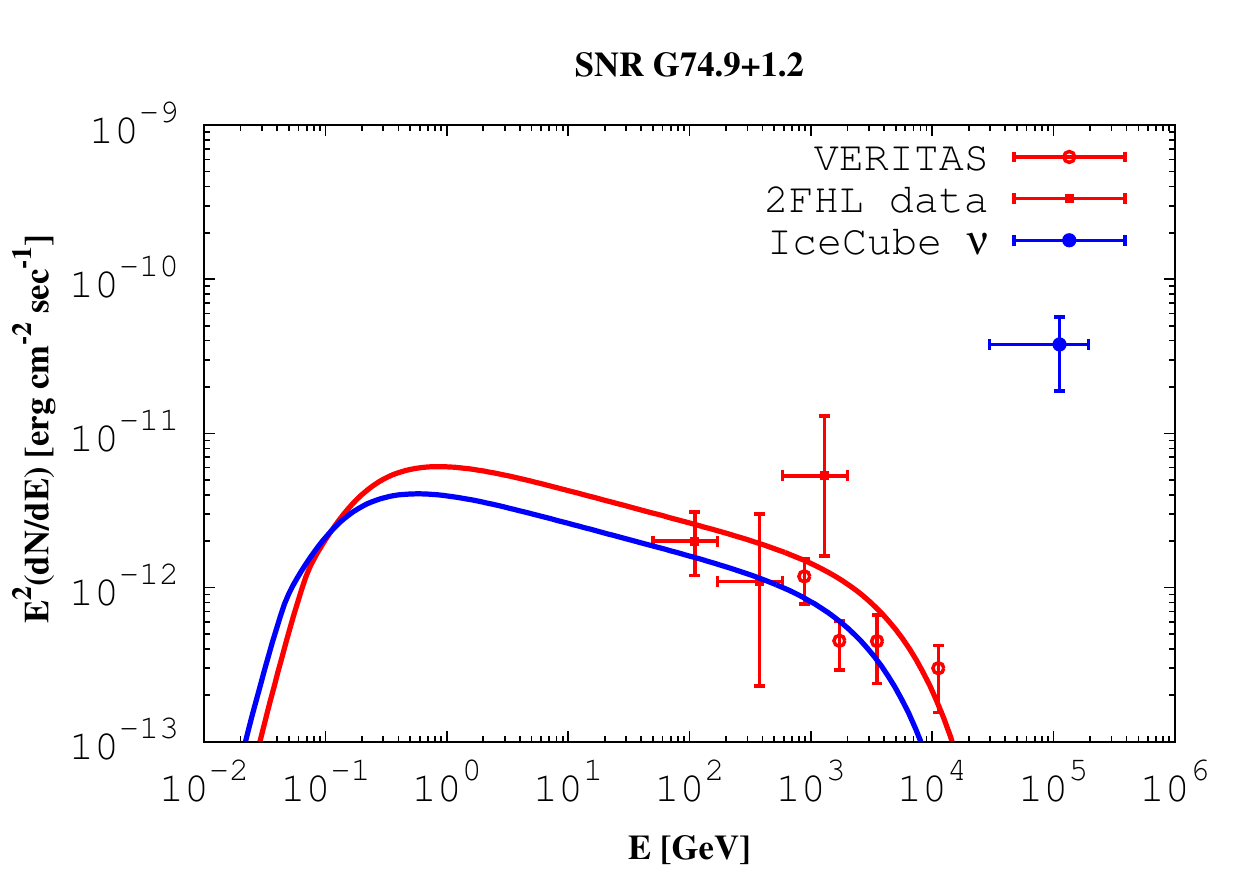}  
   
  \end{tabular}
 \caption{\label{spp_sources}The top left panel shows the $Fermi-LAT$ GeV gamma rays,  GeV-TeV gamma rays detected by H.E.S.S~\cite{tam} and the IceCube detected neutrino flux from the direction of HESSJ1614-518. The top right panel shows the TeV gamma rays detected by H.E.S.S~\cite{gchess} for the source HESS J1745-290. The bottom left panel shows the Fermi-LAT events and the H.E.S.S detected events~\cite{w41hess1,w4hess2} for W41. The bottom right panel shows the 2FHL data  and detected GeV and TeV gamma rays by VERITAS~\cite{g749} and the IceCube detected neutrino flux correlated with SNR G74.9+1.2.}
 \end{figure}

Sample-VI contains 6 sources from the 2FHL catalog which are not definitely identified but associated with SNR/PWN.  Among these 5 are correlated with the direction of the IceCube neutrino events. W31 is detected by H.E.S.S.~\cite{w4hess2}, but has not been detected with TeV events, so we have not done analysis for this source. The result of our  $pp$ interaction analysis for the 4 spp sources is shown in figure~\ref{spp_sources}.  

The result of the $pp$ interaction analysis is given in Table~\ref{startab}. The gamma ray fitting parameters, $E_0$ and $\alpha$ are listed in column 3 and 4.  The cosmic ray and gamma ray luminosity as a result of this fitting are shown in column 5 and 6, respectively. The general cosmic ray luminosity ($L_{cr}$) for sbg sources from this fitting, is within $\sim 10^{41}$ to $10^{42}$ erg/sec and the photon luminosity is within $\sim 10^{40}$ to $10^{41}$ erg/sec, which is roughly $10\%$ of the cosmic ray luminosity, as expected from the $pp$ model.  These luminosities decrease for the local starforming regions and the Galactic SNRs, however their ratio still remains at $\sim 0.1$.

With the $pp$ interaction fitting parameters we calculated the neutrino flux assuming the neutrino energy is 2/3 that of the gamma ray energy. We expected this flux to fit with the IceCube neutrino flux from the direction of the sources. However, for none of the sources the $pp$ model is able to account for the neutrino flux from the source directions inferred from the 4 year HESE sample.


\begin{table*}[tbp]\centering
\begin{adjustbox}{width=1\textwidth,center}
\ra{1.4}
\begin{tabular}{|c|c|c|c|c|c|c|}\toprule
{Sources} 
&
Neutrino ID
&
Distance
&
\multicolumn{2}{|c|}{Fitting Parameters}
&
\multicolumn{2}{|c|}{ }
\\
\hline
Names
&
Number
&
$D_L$
&
{$E_0$ [TeV]}
&
$\alpha$
&
{$L_{CR}/10^{40}$ [erg/sec]} 
& 
{$L_{\gamma}/10^{40}$ [erg/sec]}
\\
\midrule
\midrule

NGC 253                    & 7, 10, 21     & ~3.1 Mpc   & 500    & 2.4   & 6.5                  & 0.44 \\ 
NGC 1068                   & 1             & ~13.7 Mpc  & 10     & 2.5   & 115.7               & 7.61 \\ 
IC 342                     & 31             & -          & -      & -     &                      &     \\ 
M 82                       & 31            & ~3.6 Mpc   & 1000   & 2.35  & 14.6                  & 1. \\  
NGC 4945                   & 35            & ~3.9 Mpc   & 10     & 2.5   & 14                &0.61 \\ 
M 83                       & 16             & -          & -      & -     &                      & \\ 
NGC 6946                   & 34           & -          & -      & -     &                      & \\ 

\midrule
W 49A                      & 25, 34, 35    & - & -  & -   & -   & \\
Cygnus Cocoon              & 29, 34        & 50 pc      & 100    & 2.26  & $1.4 \times 10^{-7}$ & $9.5 \times 10^{-9}$\\   
30 DorC                    & 34            & 100 pc     & 1000   & 2.6   & $5 \times 10^{-8}$ & $3.4 \times 10^{-9}$\\

\midrule
\midrule

{2FHL SNRs}  &  &  &  &  &    & \\
\midrule
\midrule

SNR G315.4-2.3             & 35            & 2.5 kpc              & $100$  & 1.77   & $1.6\times 10^{-5}$  & $1.3 \times 10^{-6}$\\ 
SNR G326.3-1.8             &  -            & -                    & -      & -      & -                    & -\\ 
SNR G338.3-0.0             & 25            & 8.6 kpc              & $100 $ & 2.35   & $1.2  \times 10^{-3}$ & $8 \times 10^{-5}$\\
VelaJr                     & 40            & 0.2 kpc              & $50$   & 1.8    & $3.8 \times 10^{-7}$ & $2.8 \times 10^{-8}$\\
PuppisA                    & -             & -                    & -      & -      &    & \\
RXJ1713.7-3946             & 25            & 1 kpc                & $80$   & 1.8    & $1.15 \times 10^{-5}$ & $9 \times 10^{-7}$\\
HESSJ1800-240A             & 24, 25, 2, 14 & 2 kpc                & $100 $ & 2.6    & $9.3 \times 10^{-5}$ & $6.2 \times 10^{-6}$\\
W 28                       & 24, 25, 2, 14 & 2 kpc                & $8$    & 2.4    & $1.8 \times 10^{-4}$ & $1.2 \times 10^{-5}$\\
W 49B                     & 25, 33, 34    & 11.4 kpc   & 15       & 2.6  & $3 \times 10^{-3}$ & $3. \times 10^{-4}$\\
W 51C                      & 25, 34, 35    & 4.3 kpc              & $500 $ & 2.5    & $4.3 \times 10^{-4}$ & $2.9 \times 10^{-5}$\\
IC 443                     & -             & -                    & -      & -      &    & \\
S 147                      & -             & -                    & -      & -      &    & \\
Gamma Cygni & 29, 34        & -                    & -      & -      &    & \\
SNR G150.3+4.5             & -             & -                    & -      & -      &    & \\
Cas A                      & 34            & 3.4 kpc              & 15     & 2.1    & $5.7 \times 10^{-4}$ & $3.7 \times 10^{-5}$\\
\midrule
\midrule
{2FHL SPPs} &  &  &  &  &    &  \\
\midrule
\midrule
HESSJ1614-518              & 52            & 10 kpc               & $10 $ & 1.8    & $8.3 \times 10^{-4}$ & $5.6 \times 10^{-5}$\\
HESS J1745-290             & 25, 2, 14     & 8.5 kpc              & $250 $ & 2.37    & $1.16 \times 10^{-3}$ & $7.7 \times 10^{-5}$ \\
W 30                       & 24, 25, 2, 14 & -                    & -      & -      &    & \\ 
W 41                       & 24, 25, 2     & 4 kpc                & 10     & 2       & $1 \times 10^{-4}$ & $6.6 \times 10^{-6}$\\ 
SNR G74.9+1.2              & 29, 34        & 12 kpc               & 50     & 2.25   & $3.8 \times 10^{-3}$ & $ 2.5\times 10^{-4}$\\ 
PSR J0205+6449             & -             & -                    & -      & -      &    & \\ 
\bottomrule
\end{tabular}
\end{adjustbox}
\caption{\label{startab} 2FHL sources and the neutrino events that correlated with theses sources. For HESSJ1614-518, a spp source the luminosity distance ($D_L$) is not available so we have taken an estimated value of 10 kpc.}
\label{tab:2fhl}
\end{table*}

\section{Conclusion}
The astrophysical sources of the IceCube HESE neutrino events is one of the important puzzles since their discovery.  A report of significant correlation of the 3 year IceCube neutrino events with starforming host galaxies (sbgs) and local starforming regions~\cite{Emig:2015dma} motivated us to revisit the study with more events added in the 4th year of IceCube observation.  We carried our study with the sample sets used in~\cite{Emig:2015dma} with the cross-correlation method, and found similar result for 3 year IceCube data, shown in figure~\ref{compare}.  However, after performing the same study with 4 year data the significance increased for sbg sources but not for the TeV detected sbgs and the local starforming regions.  Moreover, we do not find any hint of correlation with the sbg sources and the local starforming regions when considering $> 60$~TeV HESE neutrino sample, which is almost free of atmospheric background.  We therefore conclude that the starburst galaxies and the local starforming regions considered in~\cite{Emig:2015dma} are most likely not the sources of the IceCube HESE neutrinos.

Furthermore we performed the cross correlation study for the highest energy gamma ray sources in the $Fermi$ 2FHL catalog that includes 33 Galactic sources such as the SNRs, PWNs which are associated to star formation.  These Galactic sources show a significant correlation with the HESE neutrinos, giving p-value (pre-trial) 0.017 for all neutrino energies and a lower p-value (pre-trial) 0.0016 for neutrinos with energy greater than 60 TeV.  A global post-trial p-value for combining the 2FHL Galactic source sample with the other star-forming sources (Samples-I and II) is reduced to $0.09-0.17$.  We have also done correlation study with the SNRs, PWNs and spp sources listed in the 2FHL Galactic sources separately but could not find a significant p-value.


For further study of the possible gamma-ray and neutrino emissions from the star-forming sources, we have done $pp$ interaction modeling for the sources that correlated with the IceCube neutrino events. We fitted the observed high energy gamma rays detected from individual sources by changing different intrinsic parameters for $pp$ interaction and then calculated the neutrino flux with those parameters. 
While $pp$ interactions can reproduce gamma-ray data in many cases, the resulting neutrino flux falls short by at least an order of magnitude from the inferred neutrino flux from the HESE sample.

Therefore, even though we found $\sim 2\sigma$ significance (post-trial) for cross correlation of the IceCube neutrino events with different starforming sources, it is still not sufficient to say that, they are the sources of the IceCube HESE sample.  A further increase in number of IceCube data can verify or discard this small significance in future.  

\section{Acknowledgment}
We thank Markus Ahlers, Claudio Koper and Cecilia Lunardini for helpful discussion. This work was supported by the National Research Foundation (South Africa), grant no. 87823 (CPRR).


\end{document}